\pdfoutput=1
\documentclass[ALICE,manyauthors]{cernphprep}
\usepackage[comma,square,numbers,sort&compress]{natbib}
\usepackage[utf8]{inputenc}
\usepackage{hyperref}
\usepackage{color}
\usepackage{lineno}
\usepackage{booktabs}
\usepackage{cernunits}
\usepackage{units} 
\usepackage[utf8]{inputenc}
\usepackage{multirow}

\newcommand{\ee}{$e^{+}e^{-}$\xspace}
\newcommand{\pgg}{$\pi^{0}\rightarrow\gamma\gamma$\xspace}
\newcommand{\egg}{$\eta\rightarrow\gamma\gamma$\xspace}
\newcommand{\pai}{\ensuremath{\pi^{0}}\xspace}
\newcommand{\e}{\ensuremath{\eta}\xspace}
\newcommand{\pT}{\ensuremath{p_{\mbox{\tiny T}}}\xspace}

\newcommand{\sNN}{$\sqrt{s_{\mbox{\tiny NN}}}$\xspace}
\newcommand{\sNNR}{$\sqrt{s_{\mbox{\tiny NN}}}$\xspace~=~200~GeV\xspace}
\newcommand{\sNNSPS}{$\sqrt{s_{\mbox{\tiny NN}}}$\xspace~=~29.1~GeV\xspace}
\newcommand{\s}{$\sqrt{s}$\xspace~=~2.76~TeV\xspace}
\newcommand{\sL}{$\sqrt{s}$\xspace~=~7~TeV\xspace}
\newcommand{\sLL}{$\sqrt{s}$\xspace~=~8~TeV\xspace}
\newcommand{\mT}{\ensuremath{m_{\mbox{\tiny T}}}\xspace}

\newcommand{\Pb}{{\mbox{Pb--Pb}}\xspace}
\newcommand{\pPb}{{\mbox{p--Pb}}\xspace}
\newcommand{\pA}{{\mbox{pA}}\xspace}
\newcommand{\AACol}{{\mbox{AA}}\xspace}

\newcommand{\dAu}{{\mbox{d--Au}}\xspace}
\newcommand{\pAu}{{\mbox{p--Au}}\xspace}
\newcommand{\pBe}{{\mbox{p--Be}}\xspace}

\newcommand{\GeVc}{GeV/$c$\xspace}
\newcommand{\dEdx}{$\mbox{d}E/\mbox{d}x$\xspace}
\newcommand{\RConv}{R_{\mbox{\tiny conv}}}

\newcommand{\RpA}{\ensuremath{R_{\mbox{\tiny pA}}}\xspace}

\newcommand{\spPb}{\sNN~=~5.02~TeV\xspace}

\newcommand{\spp}{\sNN~=~5.02~TeV\xspace}
\newcommand{\RpPb}{$R_{\rm{pPb}}$\xspace}
\newcommand{\TpPb}{\ensuremath{T_\mathrm{pPb}}\xspace}

\newcommand{\dydpt}{\ensuremath{\mathrm{d}y \mathrm{d}\pT}}
\newcommand{\elab}{$\eta_{\rm{lab}}$}
\newcommand{\sqrtS}{\ensuremath{\sqrt{s}}}

\newcommand{\Psipair}{\ensuremath{\psi_{\mbox{\scriptsize pair}}}}
\newcommand{\Xipair}{\ensuremath{\xi_{\mbox{\scriptsize pair}}}}

\newcommand{\eminus}{\ensuremath{{\rm e}^{-}}}
\newcommand{\eplus}{\ensuremath{{\rm e}^{+}}}

\begin{document}%

\begin{titlepage}
\PHyear{2018}
\PHnumber{002}      
\PHdate{10 January}   

%

\title{Neutral pion and \e meson production in \pPb collisions at \spPb}
\ShortTitle{Neutral pion and \e meson production in \pPb collisions}   

\Collaboration{ALICE Collaboration\thanks{See Appendix~\ref{app:collab} for the 
list of collaboration members}}
\ShortAuthor{ALICE Collaboration} 

\begin{abstract}
Neutral pion and $\eta$ meson invariant differential yields were measured 
in non-single diffractive \pPb collisions at \spPb
with the ALICE experiment at the CERN LHC.
The analysis combines results from three complementary photon measurements, utilizing 
the PHOS and EMCal calorimeters and the Photon Conversion Method.
The invariant differential yields of \pai and \e meson inclusive production are 
measured near mid-rapidity in a broad transverse momentum range 
of $0.3 < $~\pT $<$~\unit[20]{GeV/$c$} and $0.7 < $~\pT$ <$~\unit[20]{GeV/$c$}, respectively.
The measured $\eta$/$\pi^{0}$ ratio increases with \pT and saturates for \pT $>$ 4 \GeVc  
at 0.483~$\pm$~0.015$_{\rm stat}$~$\pm$~0.015$_{\rm sys}$. A deviation from \mT scaling is observed for \pT~$<$~\unit[2]{\GeVc}. 
The measured $\eta$/$\pi^{0}$ ratio is consistent with previous measurements from proton-nucleus and pp 
collisions over the full \pT range.
The measured $\eta$/$\pi^{0}$ ratio at high \pT also agrees within 
uncertainties with measurements from nucleus-nucleus collisions. 
The \pai and \e yields in p-Pb relative to the scaled pp interpolated reference, \RpPb, are presented for
0.3~$<$~\pT~$<$~\unit[20]{\GeVc} and 0.7~$<$~\pT~$<$~\unit[20]{\GeVc}, respectively. 
The results are compared with theoretical model calculations.
The values of \RpPb are consistent with unity for transverse momenta above \Unit{2}{\GeVc}.
These results support the interpretation that the suppressed yield of
neutral mesons measured in 
\Pb collisions at LHC energies is due to parton energy loss in the hot QCD medium.

\end{abstract}
\end{titlepage}

\setcounter{page}{2}
\section{Introduction}

Proton-nucleus (\pA) collisions are an important tool for the study of
strongly interacting matter and the Quark-Gluon Plasma (QGP), complementing and extending 
measurements carried out with high energy collisions of heavy 
nuclei \cite{Accardi:2004be}. By using a proton instead of a heavy nucleus as one of the projectiles, 
measurements of \pA collisions have unique sensitivity to the initial-state 
nuclear wave function, and can elucidate the effects of cold nuclear matter on a 
wide range of observables of the QGP \cite{Salgado:2011wc,Salgado:2016jws}. 

Measurements of inclusive distributions of hadrons at mid-rapidity at the LHC probe parton 
fractional momentum $x$ in the range $10^{-4}<x<10^{-2}$, where
nuclear modification to hadronic structure is expected to be sizable 
\cite{Salgado:2011wc}. This range extends an order of magnitude smaller in $x$ with respect to other colliders. 
Inclusive hadron measurements are also essential to constrain theoretical 
models of particle production (\cite{Albacete:2016veq} and references therein).

Within the framework of collinearly-factorized perturbative QCD (pQCD), effects 
of the nuclear environment are parameterized using nuclear-modified
parton distribution functions (nPDFs) 
\cite{Eskola:2009uj,Helenius:2012wd,Hirai:2007sx,deFlorian:2003qf,deFlorian:2011fp,Kovarik:2015cma}, 
which have been determined from global fits at next-to-leading order (NLO) to data from 
deep inelastic scattering
(DIS), Drell-Yan, and \pai production. Inclusive hadron measurements at the LHC 
provide new constraints on gluon 
nPDFs~\cite{Eskola:2009uj,deFlorian:2011fp,Deng:2010xg}, and the flavor 
dependence of sea-quark nPDFs~\cite{Armesto:2015lrg}. Hadron production 
measurements at the LHC are likewise needed to improve constraints on 
fragmentation functions (FFs) 
\cite{dEnterria:2013sgr,Sassot:2009sh,deFlorian:2014xna}.

An alternative approach to the theoretical description of hadronic structure is 
the Color Glass Condensate (CGC) \cite{Gelis:2012ri}, an effective theory for the nuclear environment 
at low $x$ where the gluon density is high and non-linear processes are expected to play a significant role. CGC-based 
calculations successfully describe measurements of particle multiplicities 
and inclusive hadron production at high \pT in pp, d$-$Au and \pPb collisions at RHIC and at the LHC 
\cite{Levin:2004ak,Albacete:2012td,Lappi:2013zma}. CGC calculations, with 
parameters fixed by fitting to DIS data, have been compared to 
particle distributions at hadron colliders, thereby testing 
the universality of the CGC description~\cite{Lappi:2013zma}. Additional measurements of 
inclusive hadron production at the LHC will 
provide new constraints on CGC calculations, and help to refine this theoretical approach.

Recent measurements of \pPb\ collisions at the LHC indicate the presence of 
collective effects in such systems, which influence inclusive hadron distributions 
\cite{CMS:2012qk,Abelev:2012ola,ABELEV:2013wsa,Abelev:2014mda,Aad:2014lta,Aaij:2015qcq,Salgado:2016jws}.
Detailed study of identified particle spectra over a broad \pT range can constrain theoretical models incorporating such effects.
For example, the EPOS3 model \cite{Werner:2013tya} requires the inclusion of 
collective radial flow in \pPb collisions to successfully describe the \pT spectrum 
of charged pions, kaons, protons, $\Lambda$ and $\Xi$ baryons
\cite{Chatrchyan:2013eya,Abelev:2013haa}. Tests of this model with neutral pions 
and \e mesons will provide additional constraints to this approach. 

The shape of the invariant production cross section of various hadron species in pp collisions 
can be approximated by a universal function of $\mT=\sqrt{\pT^2+M^2}$ (``\mT scaling'') \cite{Bourquin:1976fe} 
where $M$ is the hadron mass. 
This scaling has been tested with many different collision energies and
systems \cite{PhysRevC.75.024909,Khandai:2011cf,Adare:2010fe}, and is commonly utilized to calculate 
hadronic distributions in the absence of measurements.
Violation of \mT scaling at low \pT in pp collisions at the LHC has been 
observed for \pai and \e mesons at \sL 
\cite{Abelev:2012cn}, and at \sLL~\cite{Acharya:2017tlv}; this may arise from collective radial flow that is indicated in pp 
collisions for $\sqrt{s}>0.9$~TeV  \cite{Jiang:2013gxa}. 
However, a deviation from \mT scaling at very 
low \pT has also been observed in \pA collisions at \sNNSPS \cite{Agakishiev:1998mw}, 
where it was attributed to enhanced low \pT pion production from resonance decays.
The simultaneous measurement of \pai and \e mesons over a broad \pT range is
therefore important to explore the validity of \mT scaling in \pA
collisions. Precise measurements of
\pai and \e mesons at low \pT also provide an experimental determination of the background for 
measurements of dilepton and direct photon production \cite{Abelevetal:2014cna,Adam:2015lda}.

Strong suppression of inclusive hadron yields at high \pT has been observed in
heavy-ion collisions at RHIC \cite{Adcox:2001jp,Adler:2002xw,Adler:2003qi,Adams:2003kv,Adler:2006hu,Muller:2006ee} 
and the LHC \cite{Aamodt:2010jd,Abelev:2013kqa,Abelev:2014ypa,Roland:2014jsa,Norbeck:2014loa}. This 
suppression arises 
from partonic energy loss in the QGP \cite{Bjorken:1982tu,Wang:1991xy,Wiedemann:2009sh,Armesto:2011ht}. Measurements 
of \pPb collisions, in which the generation of a QGP over a large volume
is not expected, provide an important reference to help disentangle initial 
and final-state effects for such observables \cite{Adler:2003ii,Adams:2003im,Salgado:2016jws}. Suppression of inclusive 
hadron production is quantified by measuring \RpA, the relative rate of inclusive production in \pA compared to pp, scaled to 
account for nuclear geometry. Measurements at RHIC and at the LHC report \RpA consistent with unity 
for \pT $>$ 2 \GeVc 
\cite{Adler:2006wg,Adams:2006nd,ALICE:2012mj,Chatrchyan:2013eya,Abelev:2013haa,Aad:2016zif,Khachatryan:2016odn,Adam:2016dau}. 
Additional, 
precise measurements of the inclusive hadron production in \pPb collisions will provide a new test of this picture.

This paper presents the measurement of $\pi^0$ and $\eta$ \pT
differential invariant yields,
together with the \e/\pai ratio in non-single
diffractive (NSD) \pPb collisions at \spPb.
The measurement covers a range of $|y_{\rm lab}|<0.8$, where $y_{\rm lab}$ is the rapidity in the laboratory reference frame.
The measured
\pai spectrum is corrected for secondary neutral pions from weak decays.
The inclusive \pai and \e yield suppression (\RpPb) is determined using a pp reference that 
was obtained by interpolating previous measurements by the ALICE experiment 
of \pai and \e meson production in pp collisions at \sqrtS~=~2.76~TeV~\cite{Abelev:2014ypa,Acharya:2017hyu}, 
at 7~TeV~\cite{Abelev:2012cn}, and at 8~TeV~\cite{Acharya:2017tlv}.
The results are compared to theoretical models incorporating different approaches, including 
viscous hydrodynamics, pQCD at NLO with nuclear-modified PDFs, and a
color glass condensate model, as well as commonly used heavy-ion event generators.

The paper is organized as follows: the detectors relevant for this analysis
are described in Sect.~\ref{sec:detector}; details of the event selection are given in
Sect.~\ref{sec:event}; photon and neutral meson reconstruction, the 
systematic uncertainties as well as the calculation of the pp reference 
for the nuclear modification factor are explained in Sect.~\ref{sec:analysis};
the results and comparisons to the theoretical models are given in Sect.~\ref{sec:results} followed by
the conclusions in Sect.~\ref{sec:conclusions}.

\section{Detector description}
\label{sec:detector}

A comprehensive description of the ALICE experiment and its performance is provided 
in Refs. \cite{Aamodt:2008zz,Abelev:2014ffa}.
The \pai and \e mesons were measured via their two-photon decay channels \pgg and \egg (branching ratio 
BR = 98.823~$\pm$~0.034\% and 39.41~$\pm$~0.20\%, 
respectively), and in case of the \pai also via the Dalitz decay 
channel  $\pi^{0} \rightarrow \gamma^{*}\gamma \rightarrow e^{+}e^{-}\gamma$ 
(BR = 1.174~$\pm$~0.035\%) including a virtual photon $\gamma^{*}$ \cite{Olive:2016xmw}.
Photon reconstruction was performed in three different ways, using the electromagnetic 
calorimeters, the Photon Spectrometer (PHOS) \cite{Dellacasa:1999kd} and the Electromagnetic 
Calorimeter (EMCal) \cite{Abeysekara:2010ze}, and the photon conversion method (PCM). The PCM used 
converted \ee pairs reconstructed using charged tracks measured in the Inner Tracking 
System (ITS) \cite{Aamodt:2010aa} and the Time Projection Chamber (TPC) \cite{Alme:2010ke}.
Each method of photon and neutral meson reconstruction has its own advantages, specifically the wide 
acceptance and good momentum resolution of PCM at low \pT, and the higher \pT reach 
of the calorimeters \cite{Abelev:2012cn,Acharya:2018yhg,Abelev:2014ypa,Acharya:2017hyu}.
The combination of the different analysis methods provides independent cross-checks of the results,
a broader \pT range of the measurement, and reduced systematic and statistical uncertainties. 

The PHOS \cite{Dellacasa:1999kd} is a fine-granularity lead tungstate electromagnetic calorimeter that 
covers  $|$\elab$| < 0.12$ in the lab-frame pseudorapidity and ${260^\circ < \varphi < 320^\circ}$ in azimuth angle. 
During the LHC Run 1 it consisted of three modules at a radial distance of \unit[4.6]{m} from the ALICE interaction point. 
The PHOS modules are rectangular matrices segmented into $64 \times 56$ square cells of $2.2 \times 2.2$~cm$^2$ transverse size.
The energy resolution of the PHOS is $\sigma_E /E = 1.8\%/E \oplus  3.3\%/ \sqrt{E} \oplus 1.1\%$, with $E$ in units of GeV. 
The EMCal \cite{Abeysekara:2010ze} is a lead-scintillator sampling electromagnetic calorimeter. 
During the period in which the analyzed dataset was collected, the EMCal consisted of 10 modules
installed at a radial distance of \unit[4.28]{m} with an aperture of $|$\elab$|<0.7$ and $80^\circ < \varphi < 180^\circ$. 
The energy resolution of the EMCal is $\sigma_E /E = 4.8\%/E \oplus 11.3\%/ \sqrt{E} \oplus 1.7\%$ with 
energy $E$ in units of GeV. The EMCal modules are subdivided into $24
\times 48$ cells of $6 \times 6~\mbox{cm}^2$ transverse size.
The material budget of the active volumes of both calorimeters is about 20 radiation lenghts ($X_{\mbox{\small 0}}$).
The amount of material of the inner detectors between the interaction point and the calorimeters is about $0.2\,X_0$ for 
PHOS and ranges between $0.55\,X_0$ to $0.8\,X_0$ for EMCal, depending on the module.
The relative cell energy calibration of both calorimeters was obtained by equalization of the \pai peak position 
reconstructed in each cell with high-luminosity pp collisions.

The Inner Tracking System (ITS) consists of six layers of silicon detectors and is located directly 
around the interaction point, covering full azimuth. The two innermost layers consist of Silicon Pixel Detectors (SPD) 
positioned at radial distances 
of \unit[3.9]{~cm} and \unit[7.6]{~cm}, followed by two layers of Silicon Drift Detectors (SDD) at \unit[15.0]{~cm} 
and \unit[23.9]{~cm}, and two layers of Silicon Strip Detectors (SSD) at \unit[38.0]{~cm} and \unit[43.0]{~cm}. 
While the two SPD layers cover $|$\elab$| <$ 2 and $|$\elab$| <$ 1.4, respectively, the SDD and the SSD subtend $|$\elab$| <$ 0.9 
and $|$\elab$| <$ 1.0, respectively.
The Time Projection Chamber (TPC) is a large ($\approx$ \unit[85]{~m$^3$}) cylindrical drift detector filled  with a
Ne/CO$_2$ (90/10\%) gas mixture. It covers $|$\elab$| <$ 0.9 over the full
azimuth angle, with a maximum of 159 reconstructed space points along the track path. 
The TPC provides particle identification via the measurement of the specific energy 
loss (\dEdx) with a resolution of 5.5\%. 
The material thickness in the range $R$ $<$ \unit[180]{~cm} and $|$\elab$| <$ 0.9 amounts to (11.4 $\pm$ 0.5)\% of $X_{\mbox{\small 0}}$, 
corresponding to a conversion probability of $(8.6 \pm 0.4)$\% for high photon 
energies \cite{Abelev:2014ffa}.
Two arrays of 32-plastic scintillators, located at  $2.8 < \eta_{\rm{lab}} < 5.1$ (V0A) 
and  $-3.7 < \eta_{\rm{lab}} < -1.7$ (V0C), are used for triggering \cite{Abbas:2013taa}.

\section{Event selection}
\label{sec:event}
The results reported here use data recorded in 2013
during the LHC \pPb run at \spPb. 
Due to the 2-in-1 magnet design of the
LHC \cite{Evans:2008zzb}, which requires the same magnetic rigidity for both
colliding beams, the nucleon-nucleon center-of-mass system was moving with
$y_{\rm{NN}}=0.465$ in the direction of proton beam.
About $10^8$ \pPb collisions were recorded using a minimum-bias (MB)
trigger, which corresponds to an integrated luminosity of
$50~\mu\mbox{b}^{-1}$.
The ALICE MB trigger required a coincident signal in both 
the V0A and the V0C detectors to reduce the 
contamination from single diffractive and electromagnetic
interactions \cite{PhysRevLett.110.032301}. 

The primary vertex of the collision was 
determined using tracks reconstructed in the TPC and ITS as 
described in detail in Ref. \cite{Abelev:2014ffa}. 
From the triggered events, only events with a reconstructed vertex ($\sim$98.5\%) were 
considered for the analyses. Additionally, the $z$-position of the vertex was required to be within 
$\pm$ \unit[10]{cm} with respect to the nominal interaction point.
The event sample selected by the above-mentioned criteria mainly consisted
of non-single diffractive (NSD) collisions.
The neutral meson yields were normalized per NSD collision, which was determined from the number 
of MB events divided by the correction factor $96.4\% \pm 3.1\%$ to account for 
the trigger and vertex reconstruction efficiency
\cite{PhysRevLett.110.032301,Adam:2016dau}.
This correction factor was determined using a combination of different event generators 
and taking into account the type of collisions 
used in the analyses. This correction is based on the assumption that non-triggered events contain 
no neutral mesons at mid-rapidity; see Ref. \cite{PhysRevLett.110.032301} for details.

Pile-up events from the triggered bunch crossing, which have more than one \pPb interaction in the triggered
events, were rejected by identifying multiple collision vertices reconstructed by the SPD detector. The fraction of such
pile-up events in the analyzed data sample was at the level of 0.3\%.

\section{Data analysis}
\label{sec:analysis}

\subsection{Photon and primary electron reconstruction}
\label{sec:PhoElec}

Photons and electrons hitting the PHOS or the EMCal produce
electromagnetic showers which deposit energy in multiple cells. 
Adjacent fired cells with energies above $E^{\rm min}_{\rm cell}$ were grouped together into clusters. 
Noisy and dead channels were removed from the analysis prior to clusterization.
The clusterization process started from cells with an energy exceeding $E_{\rm seed}$.
The choice of the values of $E_{\rm seed}$ and 
$E^{\rm min}_{\rm cell}$ was driven by the energy deposited by a minimum ionizing particle, 
the energy resolution, noise of the electronics, and optimizing the signal to background ratio of meson candidates. 
For PHOS, $E_{\rm seed} = 50$~MeV and $E^{\rm min}_{\rm cell} = 15$~MeV were chosen. The corresponding thresholds for EMCal were
$E_{\rm seed} = 500$~MeV and $E^{\rm min}_{\rm cell} = 100$~MeV.
The photon reconstruction algorithm in PHOS separates the clusters produced
by overlapping showers from close particle hits, via a cluster unfolding procedure.
Due to a low hit occupancy in the calorimeters in \pPb collisions, relatively
loose selection criteria were applied for clusters to maximize the
neutral meson reconstruction efficiency and minimize systematic
uncertainties from photon identification criteria. 
The minimum number of cells in a
cluster was set to three and two for PHOS and EMCal, respectively, to
reduce contributions of non-photonic clusters and noise. 
Consequently, the energy threshold for PHOS and EMCal clusters was set to \unit[0.3]{GeV} and \unit[0.7]{GeV}, 
respectively.

Apart from the cluster selection criteria described above, additional detector-specific criteria
were applied in the PHOS and EMCal analyses to increase the purity and signal to background ratio of the photon sample. 
The EMCal clusters were selected in $|$\elab$| <0.67$ and $80^\circ<\varphi<180^\circ$, which is the full EMCal acceptance during the LHC Run 1 p-Pb run.
In the EMCal analysis, the purity of the photon sample was enhanced by
rejecting charged tracks reconstructed in the TPC that are matched to
a cluster in the EMCal. 
The matching criteria, based on the distance between the track and the cluster in $\eta$ and $\varphi$, 
depend on the track \pT to maximize purity at low \pT and statistics at high \pT.
The purity is further enhanced by requirements on
the squared major axis of the cluster shape $\sigma_{\rm long}^{2}$
calculated as the principle eigenvalue of the cluster covariance matrix $s_{ij}$ via
$\sigma_{\rm long}^{2} =(s_{\eta\eta}+s_{\varphi\varphi})/2 
+\sqrt{(s_{\eta\eta}-s_{\varphi\varphi})^{2}/4+s^{2}_{\eta\varphi}}$
where $s_{ij} = \langle ij \rangle - \langle i \rangle\langle j \rangle$
are the covariance matrix elements, $i,j$ are cell indices in $\eta$ or $\varphi$ axes respectively,
$\langle ij \rangle$ and $\langle i \rangle$, $\langle j \rangle$ are
the second and the first moments of the cluster cells weighted with
the cell energy
logarithm~\cite{Alessandro:2006yt,Acharya:2017hyu}. Photon clusters in
EMCal and PHOS were defined by the condition $0.1 < \sigma_{\rm long}^2 < 0.5$ and $\sigma_{\rm long}^2 > 0.2$, respectively, which selected clusters with axial symmetry.

In addition to these requirements, a selection criterion on cluster timing was applied in order
to exclude clusters from other bunch crossings. 
Since the minimum interval between colliding bunches was \unit[200]{ns}, 
$|t|<100$~ns had to be fulfilled for PHOS.
For EMCal the cell time of the leading cell of the cluster 
was required to be within $|t|<50$~ns of the time of the triggered bunch crossing.

Photons converted into \ee pairs were reconstructed with 
a secondary-vertex algorithm that searches for oppositely-charged track pairs
originating from a common vertex, referred to as V$^0$ \cite{Abelev:2014ffa}.
Three different types of selection criteria were 
applied for the photon reconstruction: requirements on the charged track quality, particle identification criteria 
for electron selection and pion rejection, 
and requirements on the V$^{0}$ sample that exploit the specific topology of a photon conversion.
Details of the PCM analysis and the selection criteria are described in Refs.~\cite{Abelev:2012cn,Abelev:2014ypa}.
Electron identification and pion rejection were performed by using the specific energy loss \dEdx in the TPC. 
Detailed requirements are listed in \hyperref[tab:CutValuesPCM]{Table~\ref*{tab:CutValuesPCM}}, 
where $n\sigma_{e}$ and  $n \sigma_{\pi}$ are deviations of \dEdx from the electron and pion expectation 
expressed in units of the standard deviation $\sigma_{e}$ and $\sigma_{\pi}$, respectively.
In comparison to the previous analyses of the $\gamma\gamma$ decay channel ({PCM$-\gamma \gamma$) 
\cite{Abelev:2012cn,Abelev:2014ypa}, the converted photon topology selection criteria were slightly 
modified to further increase the purity of the photon sample.
The constant selection criterion on the $e^\pm$ transverse momentum with respect to the V$^0$ 
momentum, $q_{\mbox{\tiny T}}$, was replaced by a two-dimensional selection in the ($\alpha$,$q_{\mbox{\tiny T}}$) distribution, 
known as the Armenteros-Podolanski plot \cite{podolanski1954iii},
where $\alpha$ is the longitudinal momentum asymmetry of positive and negative tracks, 
defined as $\alpha = (p_L^+ - p_L^-)/(p_L^+ + p_L^-)$.
The fixed selection criterion on the reduced $\chi^{2}$ of the converted photon fit
to the reconstructed V$^0$ was changed to 
the $\psi_{\rm pair}$-dependent $\chi^{2}$ selection, where
$\psi_{\rm pair}$ is the angle between the plane that is perpendicular to the magnetic 
field ($x$-$y$ plane) and the plane defined by the opening angle of the pair \cite{Dahms:2006zg}. 
It is defined as $\Psipair= \mbox{arcsin} \left({{\Delta \theta}\over{\Xipair}}\right)$, 
where $\Delta \theta $ is the polar angle difference between electron and positron tracks, 
$\Delta \theta = \theta(\eplus)-\theta(\eminus)$, and 
$\Xipair$ is the total opening angle between them. 
For converted photons with vanishing opening angle between the \ee pair the \Psipair\ distribution is peaked at zero,
while it has larger or random values for virtual photons of the Dalitz decay or combinatorial background, respectively.
The applied selection criteria on the converted photon for the PCM-$\gamma\gamma$ and PCM-$\gamma^*\gamma$ decay channels are summarized 
in \hyperref[tab:CutValuesPCM]{Table~\ref*{tab:CutValuesPCM}}.
\begin{table}[ht]
\centering 
\begin{tabular}{|p{4.3cm}c|c|}
  \cline{2-3}
  \multicolumn{1}{c|}{} & \multicolumn{1}{c|}{\textbf{PCM$-\gamma \gamma$}} & \multicolumn{1}{c|}{\textbf{PCM$-\gamma^{*} \gamma$}} \\ \hline
  \multicolumn{1}{|l|}{\textbf{Track reconstruction }} & \multicolumn{2}{c|}{} \\
  $e^\pm$ track \pT       & \multicolumn{2}{|c|}{$\pT > 0.05$~\GeVc}  \\ 
  $e^\pm$ track $\eta$    & \multicolumn{2}{|c|}{$|\eta_{\rm{lab}}|<0.9$} \\ 
  $N_{\mbox{\tiny clusters}}/N_{\mbox{\tiny findable clusters}}$ &  \multicolumn{2}{|c|}{$>60\%$} \\ 
  conversion radius      & \multicolumn{2}{|c|}{$5 <\RConv<180$~cm} \\ \hline
  \multicolumn{1}{|l|}{\textbf{Track identification }} & \multicolumn{1}{c|}{} & \multicolumn{1}{c|}{} \\
  $n \sigma_{e}$ TPC  & \multicolumn{1}{|c|}{$-4  < n \sigma_{e} <5$}  	 & $-4  < n \sigma_{e} <5$ 		                \\ 
  \multirow{2}{*}{$n \sigma_{\pi}$ TPC}      & \multicolumn{1}{|c|}{$n \sigma_{\pi} >1$ at $0.4<p<100$~GeV/$c$} & $n \sigma_{\pi} >2$ at $0.5<p<3.5$~GeV/$c$ \\
      & \multicolumn{1}{|c|}{}	& $n \sigma_{\pi} >0.5$ at $p>3.5$~GeV/$c$	\\ \hline
  \multicolumn{1}{|l|}{\textbf{Conversion $\gamma$ topology }} & \multicolumn{1}{c|}{} & \multicolumn{1}{c|}{} \\
 $q_{\mbox{\tiny T}}$        & \multicolumn{1}{|c|}{\strut $q_{\mbox{\tiny T}} < 0.05\sqrt{1-(\alpha/0.95)^2}$~\GeVc}& 	$q_{\mbox{\tiny T}}<$ \unit[0.15]{GeV/$c$} \\ 
  photon fit quality     & \multicolumn{1}{|c|}{$\chi^2_{\mbox{\tiny max}} = 30$}    &  $\chi^2_{\mbox{\tiny max}} = 30$ \\ 
  $\psi_{\rm pair}$  &  \multicolumn{1}{|c|}{$|\psi_{\rm pair}|< 0.1\,(1 - \chi^2 / \chi^2_{\mbox{\tiny max}})$} & --- 		\\ \hline
 \end{tabular}
  \caption{Selection criteria of the converted photon reconstruction with PCM for the two-photon (PCM$-\gamma \gamma$) and the Dalitz decay channel (PCM$-\gamma^{*} \gamma$).}
  \label{tab:CutValuesPCM}
\end{table}

Virtual photons ($\gamma^{*}$) of the Dalitz decays were reconstructed from primary electrons
and positrons with the ITS and the TPC for transverse momenta $\pT>0.125$~\GeVc. 
Tracks were required to 
cross at least 70 TPC pad rows, with the
number of TPC clusters to be at least 80\% of the number expected from the geometry of the track's trajectory in the detector.
Track selection was based on $\chi^{2}$ of the ITS and TPC clusters fit to the track. 
To ensure that the selected tracks came from the primary vertex, their distance
of closest approach to the primary vertex in the longitudinal direction (DCA$_{z}$) was required to be smaller than \unit[2]{cm}
and  DCA$_{xy} <0.0182$~cm $+0.0350$~cm$/\pT^{1.01}$ in the transverse plane with \pT~given in GeV/$c$ which correspond 
to a 7 $\sigma$ selection \cite{Abelev:2014ffa}.
In addition, in order to minimize the contribution from photon conversions in the beam pipe and part of the SPD, 
only tracks with at least one hit in any layer of the SPD were accepted.
Electrons were identified by the TPC \dEdx by requiring that tracks fall within $-4 < n \sigma_{e} <5$ of the electron hypothesis.
For the pion rejection at intermediate \pT the same $n \sigma_{\pi}$ selection as described for the conversion electron 
tracks was used while at high \pT the selection was not applied, to increase the efficiency.

For the neutral meson reconstruction via the Dalitz decay channel a $\gamma^{*}$ is constructed from the primary $e^{+}e^{-}$ 
pairs and is treated as real $\gamma$ in the analysis, except with non-zero mass.
The pion contamination in the primary electron sample was reduced by constraints on the 
$\gamma^{*}$ invariant mass ($M_{\gamma^{*}} < 0.015$~GeV/$c^{2}$ at $\pT < 1$~GeV/$c$ 
and $M_{\gamma^{*}} < 0.035$~GeV/$c^{2}$ at $\pT > 1$~GeV/$c$) exploiting that most of the $\gamma^{*}$ from $\pi^{0}$ Dalitz decays 
have a very small invariant mass, as given by the Kroll-Wada formula \cite{KrollWada}.
Contamination of the $\gamma^{*}$ sample by $\gamma$ conversions was suppressed by requiring the
primary \ee pairs to satisfy $|\psi_{\rm pair}|<0.6 - 5\Delta\varphi$ and $0<\Delta\varphi<0.12$, where 
$\Delta\varphi=\varphi(e^{+})-\varphi(e^{-})$ is the difference between electron and positron azimuth angles.

\subsection{Meson reconstruction}
\label{sec:Meson}
The \pai and \e meson reconstruction was done by pairing $\gamma\gamma$ or $\gamma^{*}\gamma$ candidates and 
calculating their invariant mass in transverse momentum intervals. For simplicity, the notation PCM-EMC will stand for the 
method with one photon reconstructed via PCM and the second photon reconstructed in EMCal. PCM, EMC and PHOS refer to the
 methods with both photons reconstructed by the same methods. PCM-$\gamma^{*}\gamma$ is the method of 
meson reconstruction via the Dalitz decay channel.
In total, five different measurements (PCM, 
PCM-$\gamma^{*}\gamma$, EMC, PCM-EMC and PHOS) were done for the \pai meson and three
different ones (PCM, EMC and PCM-EMC) for the \e meson. The reconstruction of \e mesons is not accessible by PHOS due to 
the limited detector acceptance and the wider opening angle of the decay photons compared to the \pai.

Examples of invariant mass distributions are shown in 
\hyperref[fig:Pi0InvMass]{Fig.~\ref*{fig:Pi0InvMass}} and 
\hyperref[fig:EtaInvMass]{Fig.~\ref*{fig:EtaInvMass}} for selected \pT intervals for \pai and \e mesons, respectively. 
The combinatorial background, estimated using the event mixing technique \cite{Kopylov:1974th}, was scaled 
to match the background outside the signal region and subtracted from the total signal. The shape of the combinatorial 
background was optimized by mixing events within classes of similar primary vertex position
and for all methods except PHOS also similar photon multiplicity.
In case of the EMC analysis a minimal opening angle selection between the two photons of 17~mrad between the cluster 
seed cells was applied, which corresponds to 1 cell diagonal at mid rapidity, in order to provide a good event mixed 
background description. For PCM and PCM-EMC an opening angle selection of 5~mrad was applied.
The background-subtracted signal was fitted to reconstruct the mass position $(M_{\pai,\e})$ and width of 
the \pai and \e mesons. In case of the 
PCM, PCM-$\gamma^{*}\gamma$, EMC, and PCM-EMC analyses, the fit function consisted of a Gaussian 
function convolved with an exponential low-energy 
tail to account for electron bremsstrahlung \cite{Matulewicz1990194} and an additional linear function to take into 
account any residual background. 
For the PHOS analysis a Gaussian function was used.

\begin{figure}[htbp]
  \includegraphics[width=0.48\textwidth]{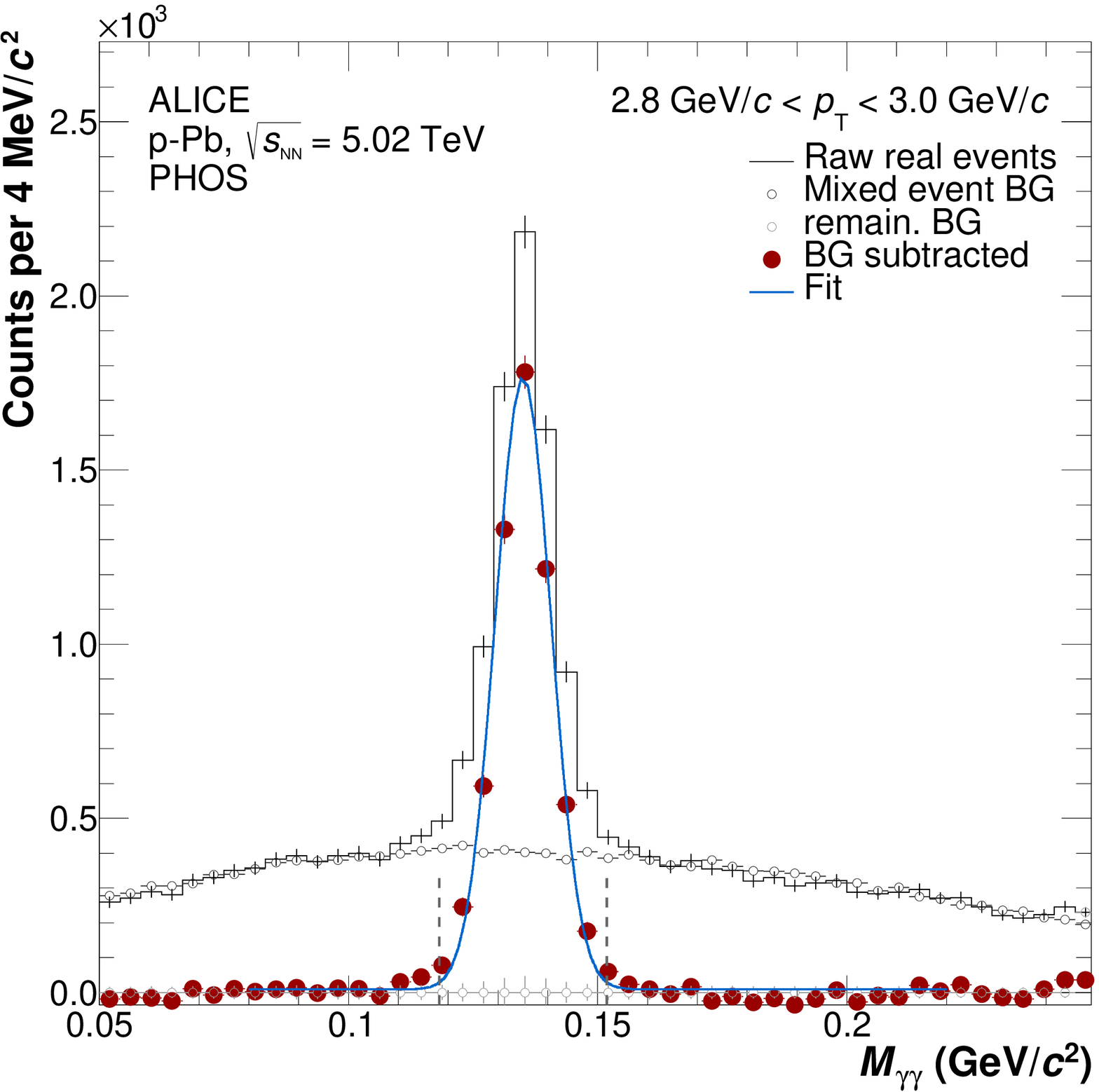}
  \hfil
  \includegraphics[width=0.48\textwidth]{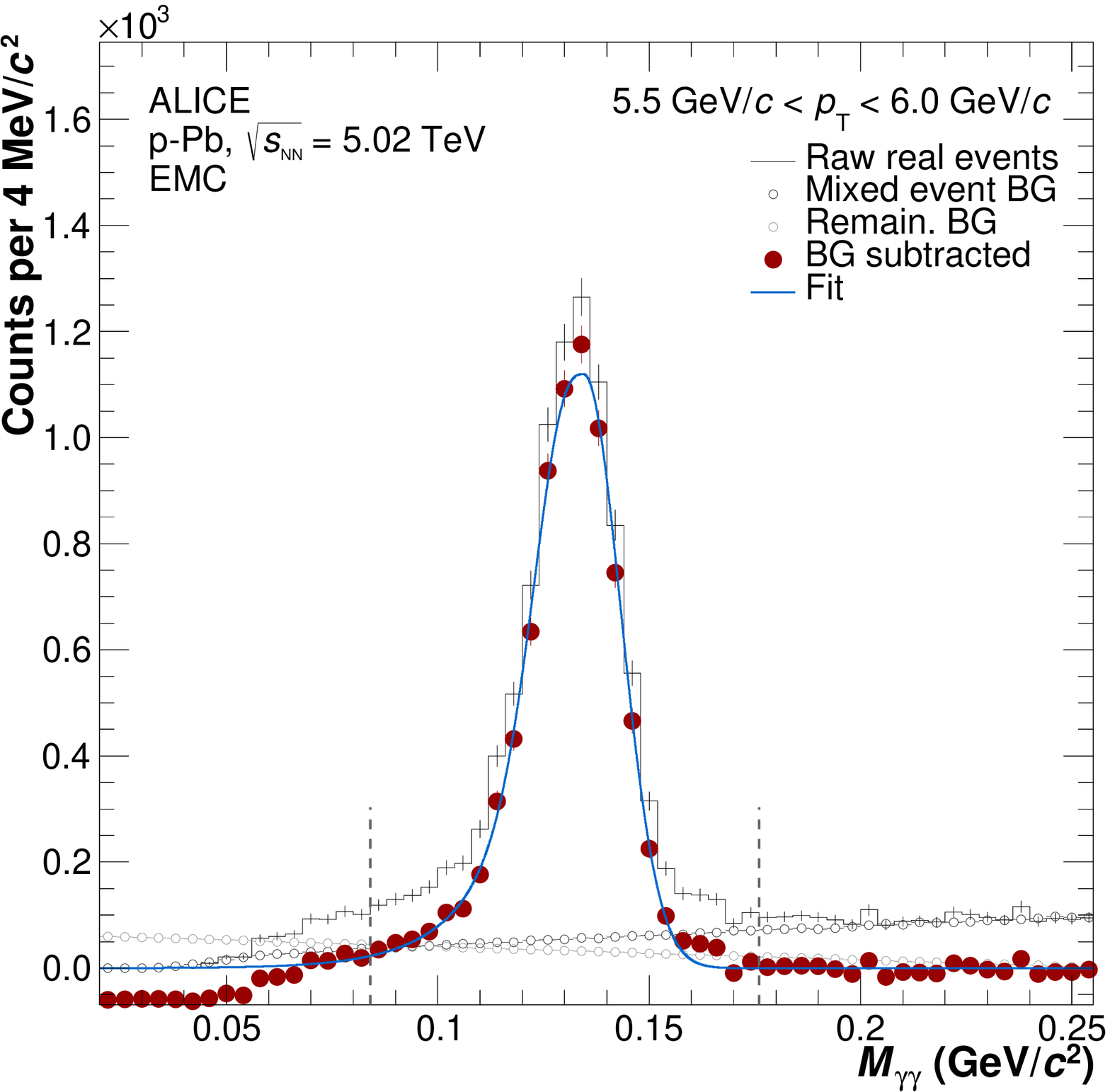}\\
  \includegraphics[width=0.48\textwidth]{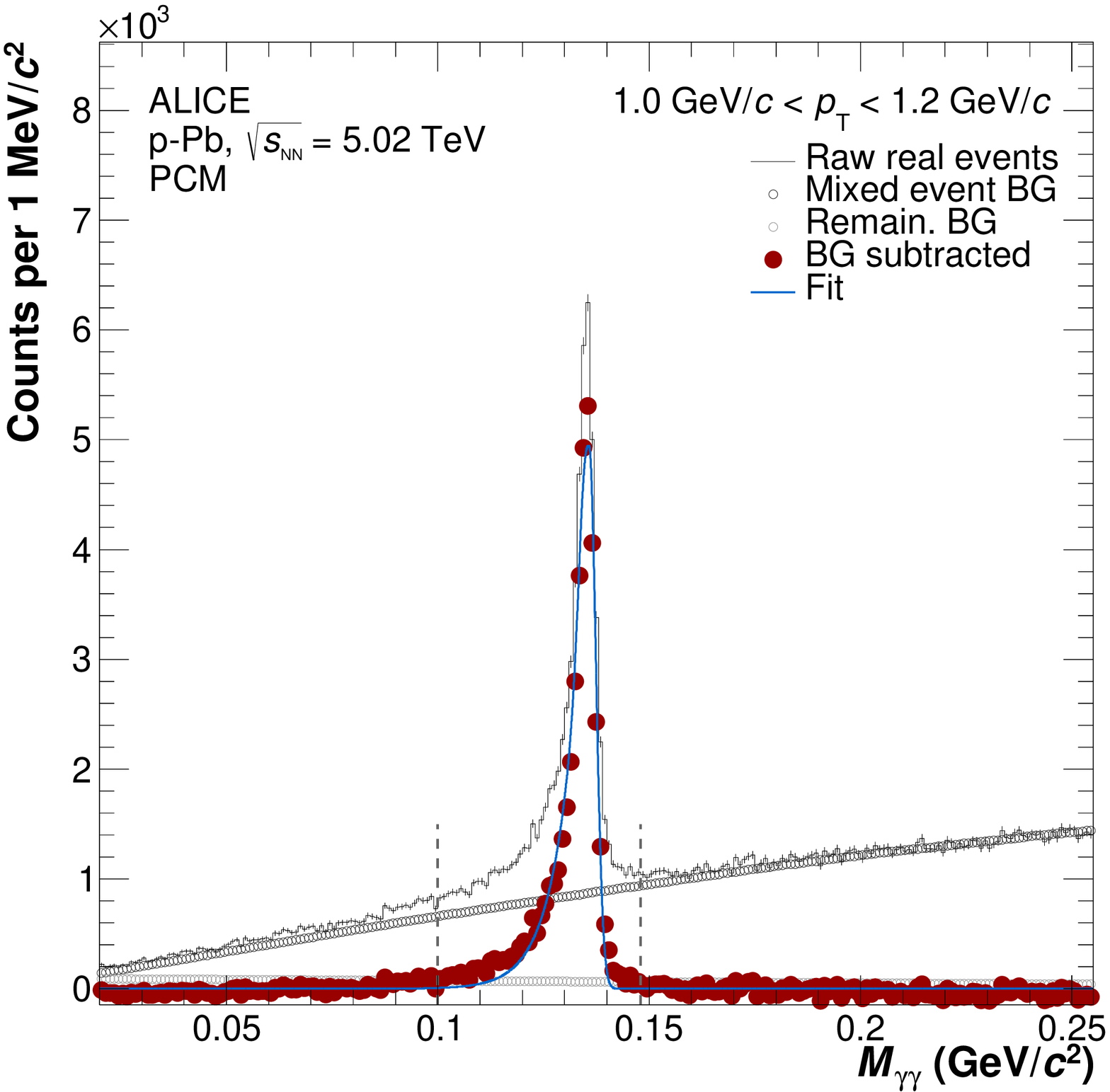}
  \hfil
 \includegraphics[width=0.48\textwidth]{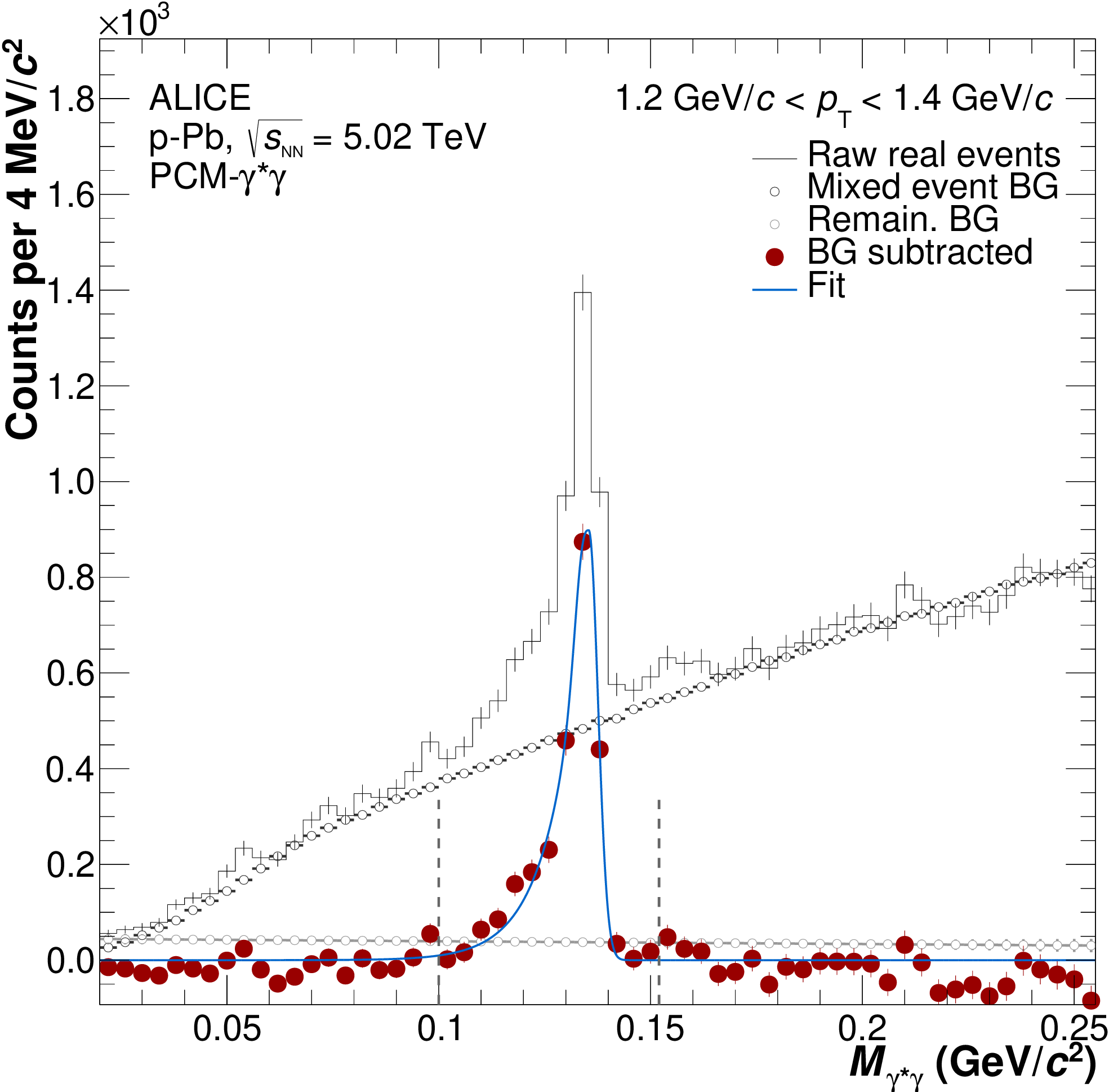}\\
  \hfil
  \includegraphics[width=0.48\textwidth]{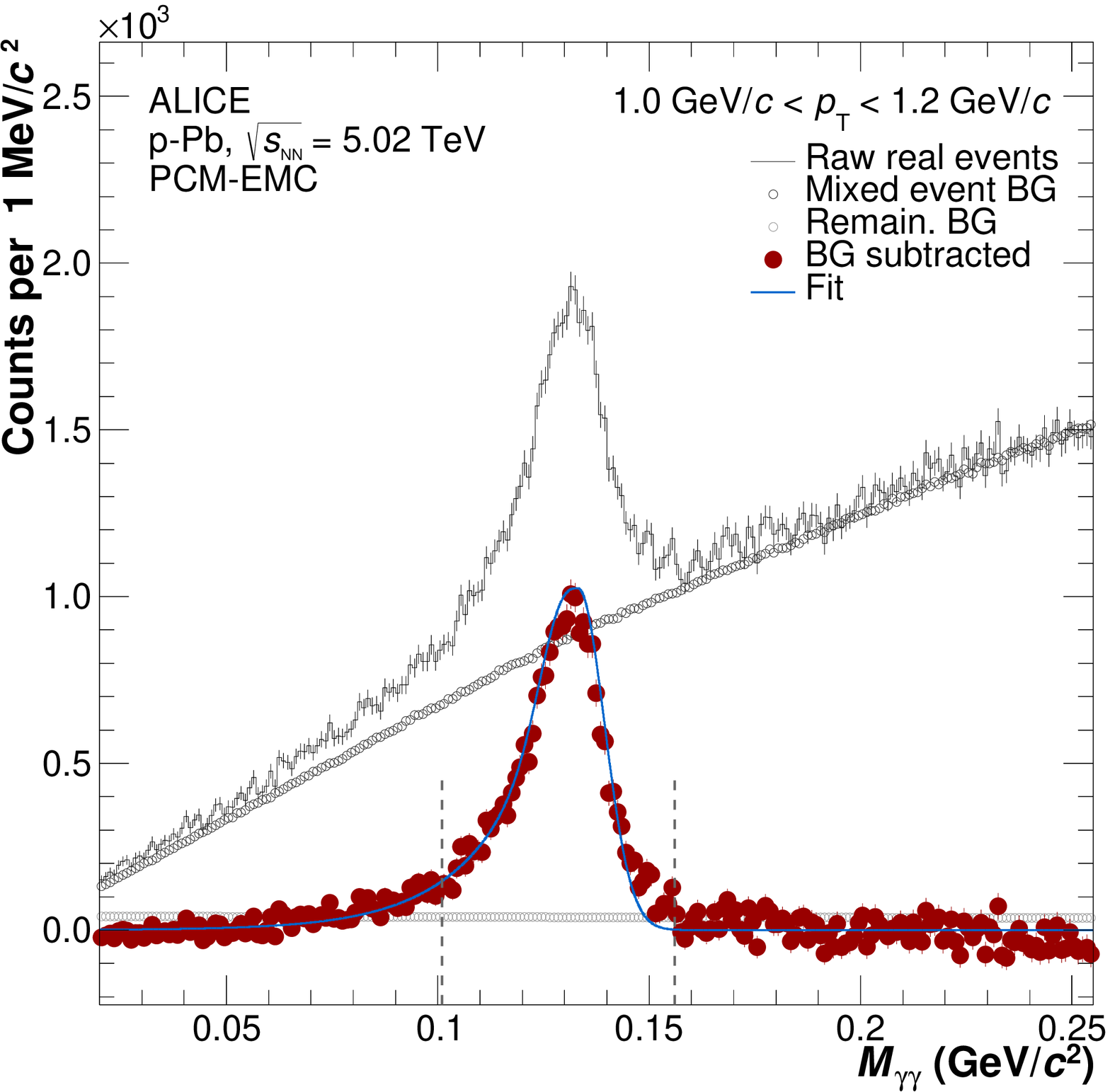}
  \hfil
  \caption{The diphoton invariant mass distributions around the \pai mass for selected intervals in \pT, without and with
combinatorial background for each of the five measurements: PHOS, EMC, PCM, PCM-$\gamma^{*}\gamma$, and PCM-EMC. 
The vertical lines correspond to the limits of the region used to compute the integration of the meson signal.}
  \label{fig:Pi0InvMass}		
\end{figure}

\begin{figure}[htb]
  \includegraphics[width=0.48\textwidth]{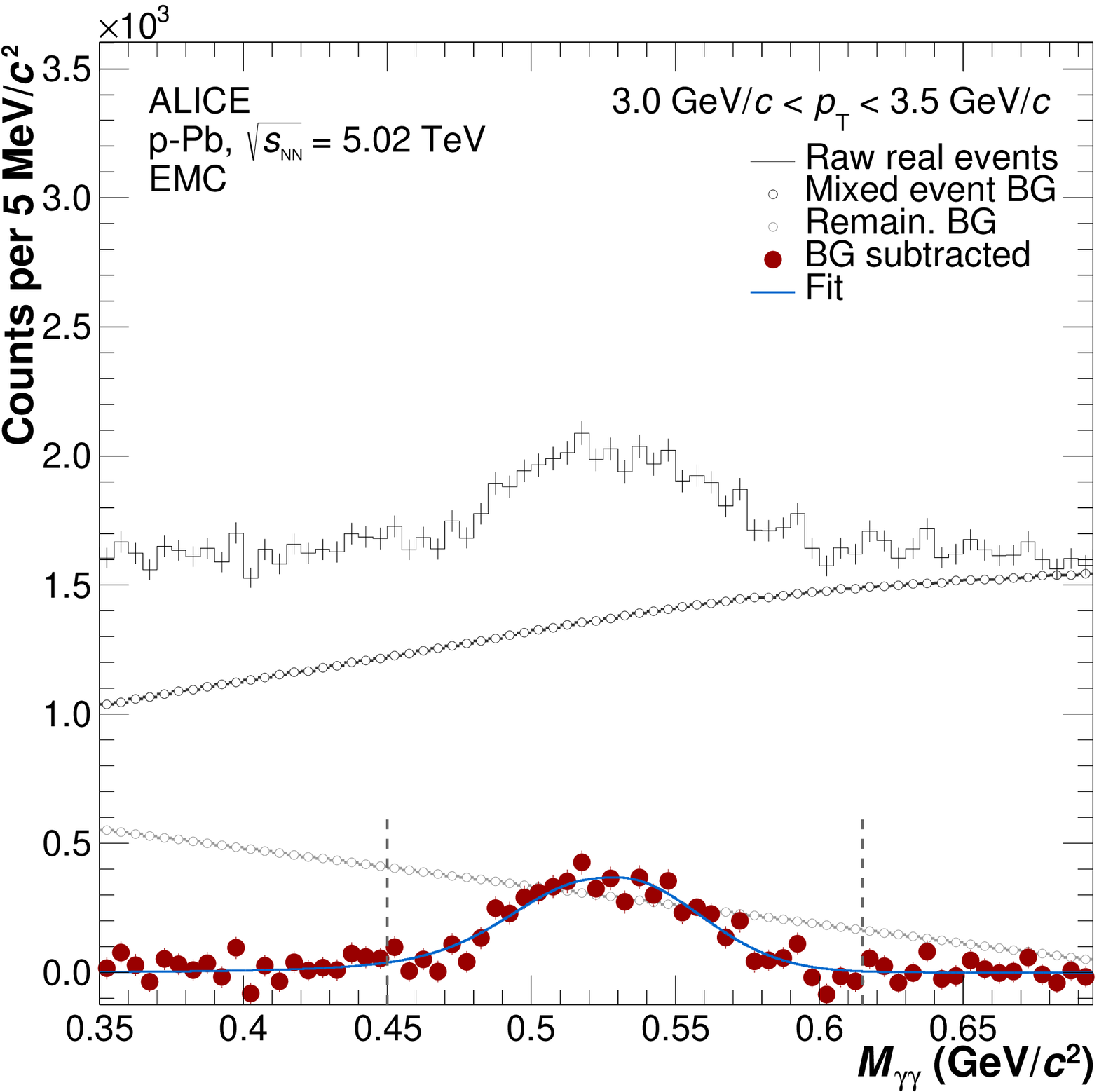}
  \hfil
  \includegraphics[width=0.48\textwidth]{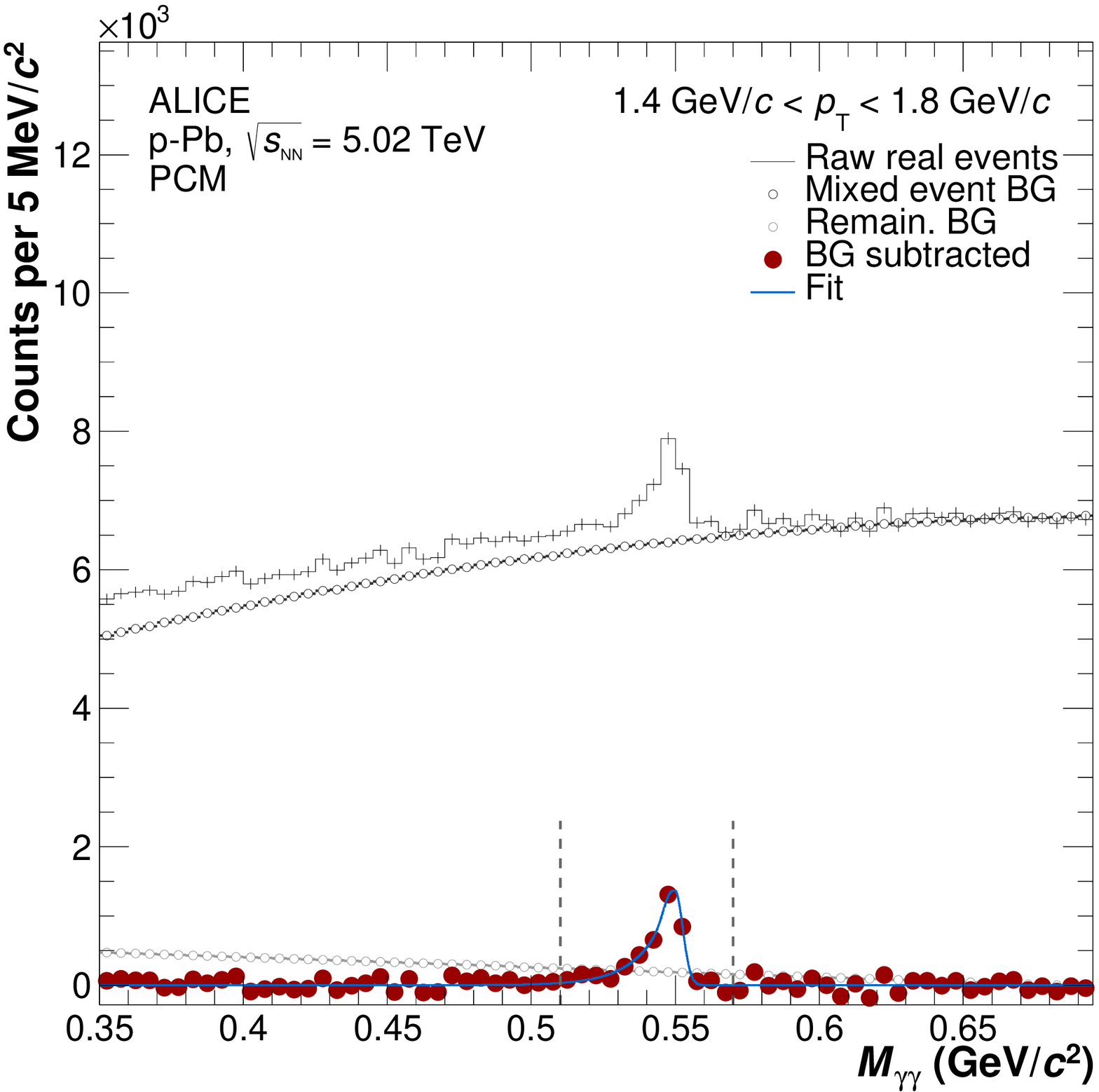}\\
  \hfil
  \includegraphics[width=0.48\textwidth]{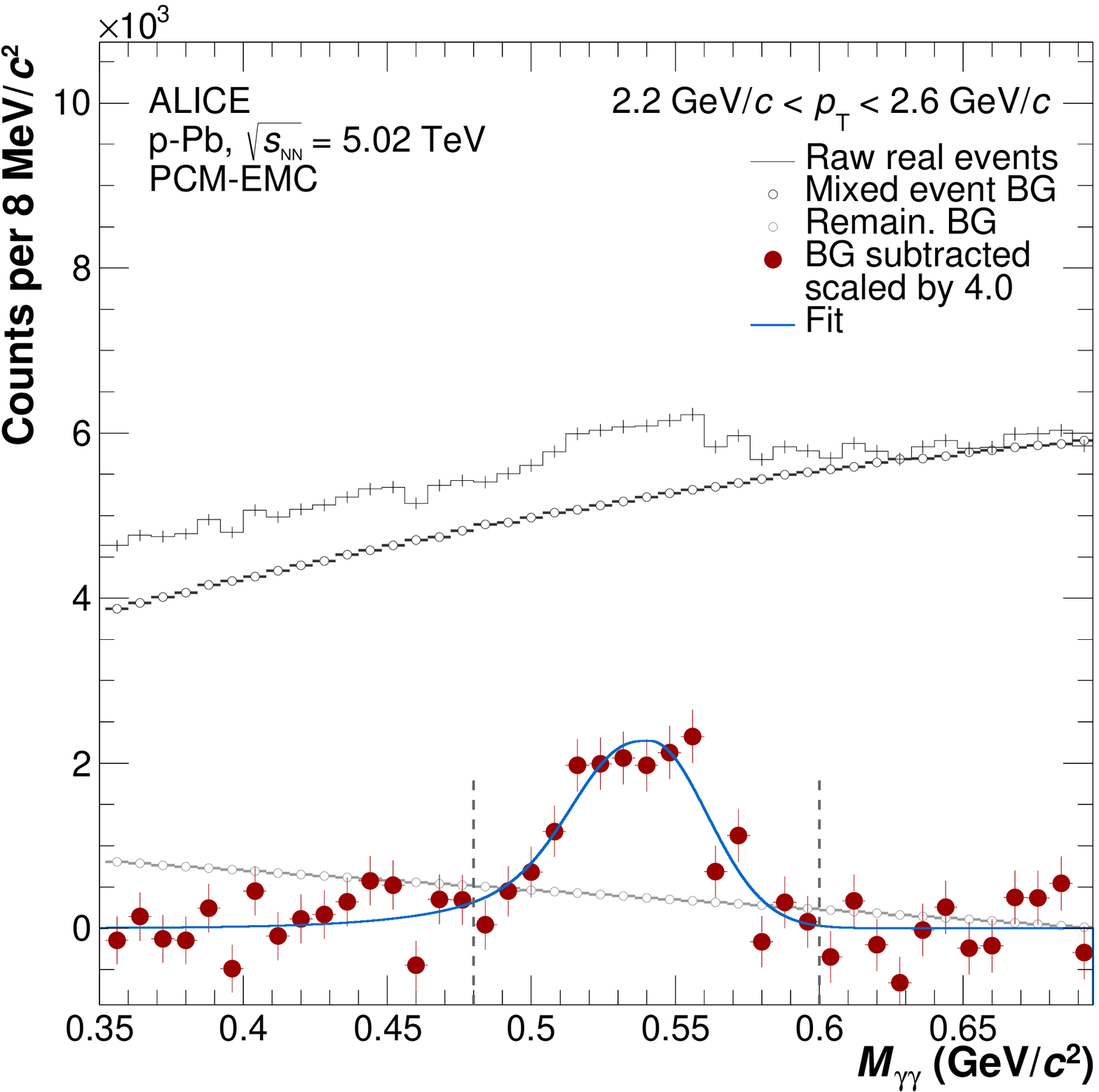}
  \hfil
  \caption{The diphoton invariant mass distribution around the \e mass for selected intervals in \pT, 
without and with combinatorial background for each of the three measurements: EMC, PCM, and PCM-EMC. 
The vertical lines correspond to the limits of the region used to compute the integration of the meson signal.}
  \label{fig:EtaInvMass}		
\end{figure}
The reconstructed \pai and \e meson peak position and width 
versus \pT compared to GEANT3 \cite{Brun:1119728} simulations are shown 
in \hyperref[fig:Pi0MassWidth]{Fig.~\ref*{fig:Pi0MassWidth}} and \hyperref[fig:EtaMassWidth]{Fig.~\ref*{fig:EtaMassWidth}}, respectively. 
The reconstructed meson mass peak position and width for each method are in good agreement 
for data and MC. The \pai and \e meson peak position for EMC and PCM-EMC was not calibrated 
to the absolute meson mass, but the cluster energy in MC was corrected by a \pT dependent correction 
factor such that the \pai mass peak positions in data and MC match within $0.4$\% for EMC and $0.5$\% for PCM-EMC. 
The cluster energy correction factor was calculated with \pai mesons reconstructed with the PCM-EMC method 
where the energy resolution of converted photons is much better than the one of real photons detected in EMC. 
Deviations of the MC \pai peak position with respect to the measured one in data were fully assigned to the EMC cluster
energy. The \pai mass peak positions in PHOS were also tuned in MC to achieve
a good agreement with data, which was done with a cluster energy correction.

The \pai and \e raw yields were obtained by integrating the background-subtracted $\gamma\gamma$ or $\gamma^{*}\gamma$ 
invariant mass distribution. The integration window around the reconstructed peak of the meson mass was determined by the 
fit function. The integration ranges, as shown in \hyperref[tab:IntWindows]{Table~\ref*{tab:IntWindows}}, were selected 
according to the resolution of respective methods. 
\begin{figure}[htb]
  \parbox[t]{0.49\hsize}{
    \includegraphics[width=\hsize]{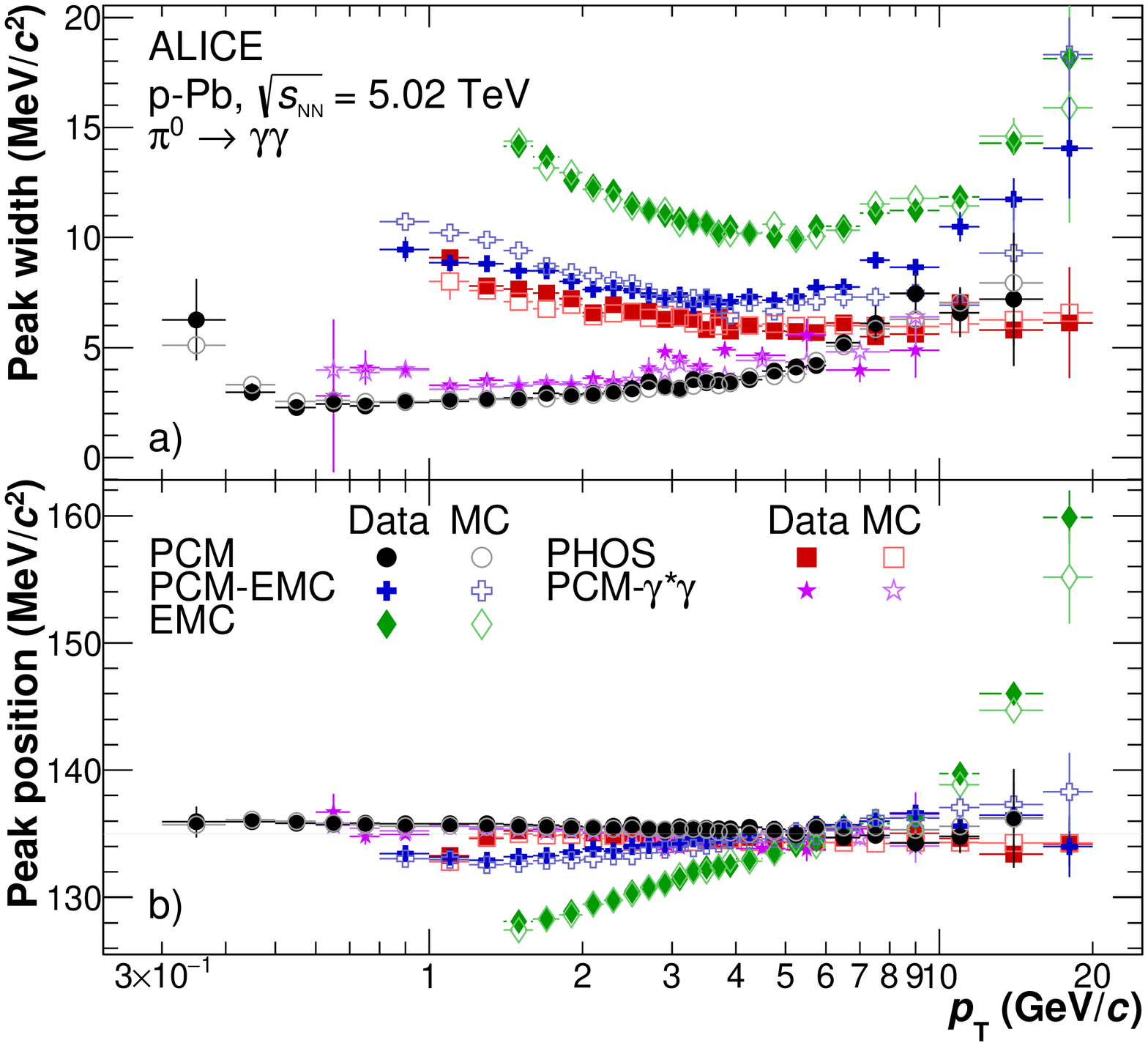}
    \caption{Reconstructed \pai mass width (top) and position (bottom) versus \pT for data and Monte Carlo simulation for all five methods.}
    \label{fig:Pi0MassWidth}
  }
  \hfil
  \parbox[t]{0.49\hsize}{
    \includegraphics[width=\hsize]{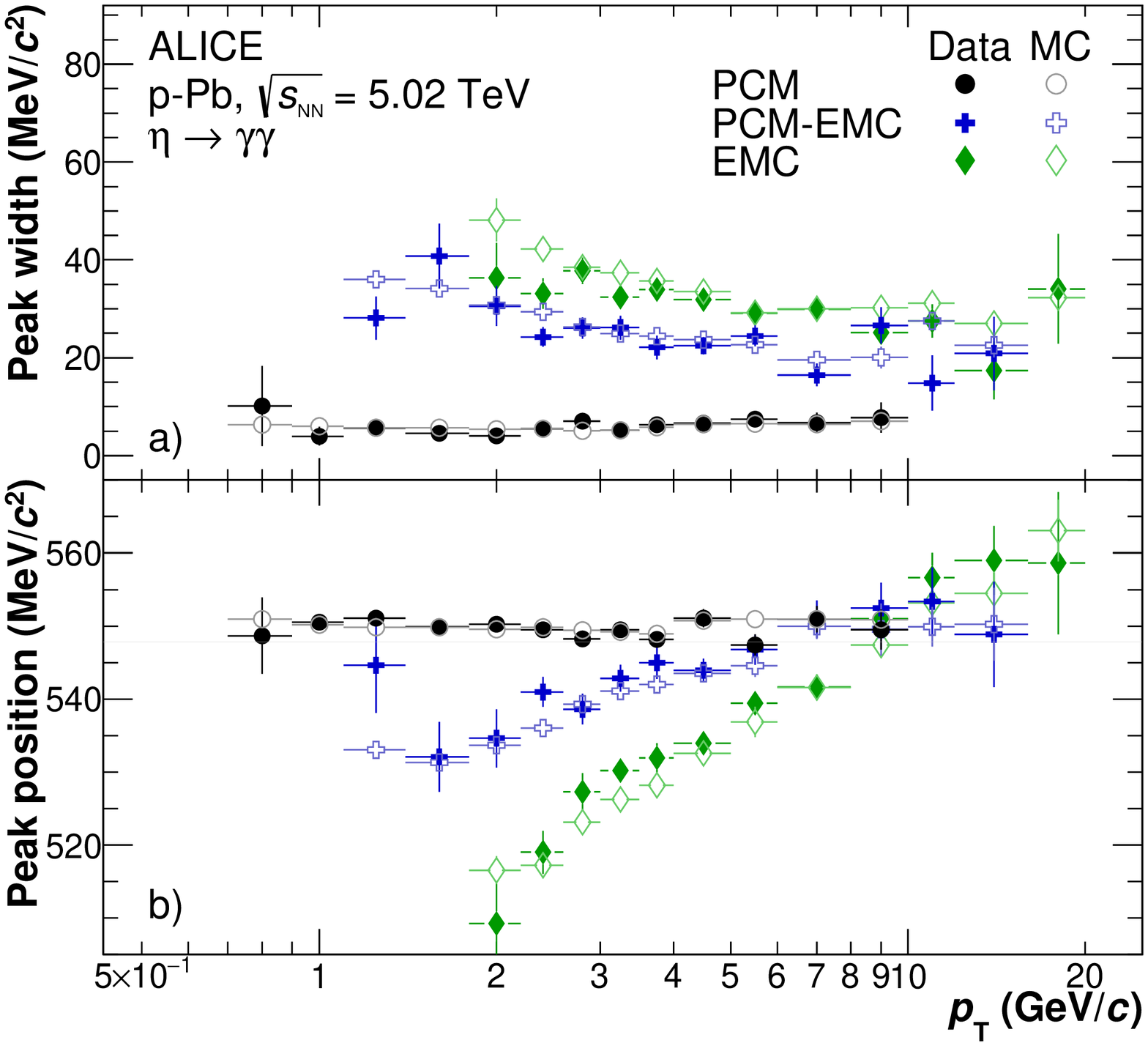}
    \caption{Reconstructed \e mass width (top) and position (bottom) versus \pT for data and Monte Carlo simulation for PCM, EMC and PCM-EMC.}
    \label{fig:EtaMassWidth}
  }
\end{figure}
\begin{table}[htb]
\small
\centering 
\renewcommand{\arraystretch}{1.15}
  \begin{tabular}{p{1.8cm}|c|c|}
  \cline{2-3}
                 & $M-M_{\pai}~(\mbox{GeV}/c^{2})$ & $M-M_{\e}~(\mbox{GeV}/c^{2})$		\\ \hline
  \multicolumn{1}{|p{1.8cm}|}{  \textbf{PHOS} }          			& $[-3\sigma,+3\sigma]$ & 	\\ 
  \multicolumn{1}{|p{1.8cm}|}{  \textbf{EMC} }          				& $[-0.05,+0.04]$ & $[-0.08,+0.08]$	\\ 
  \multicolumn{1}{|p{1.8cm}|}{  \textbf{PCM} }          				& $[-0.035,+0.01]$ & $[-0.048,+0.022]$	\\ 
  \multicolumn{1}{|p{1.8cm}|}{  \textbf{PCM$-\gamma^{*}\gamma$} }    & $[-0.035,+0.01]$ & 	\\ 
  \multicolumn{1}{|p{1.8cm}|}{  \textbf{PCM-EMC} } 					& $[-0.032,+0.022]$ & $[-0.06,+0.055]$	\\ \hline
  \end{tabular}
  \caption{Integration windows for the \pai and \e meson invariant mass distributions, where $M_{\pai}$ and $M_{\eta}$ are the reconstructed 
mass positions from the fit, and $M$ is the nominal mass of the respective meson.}
  \label{tab:IntWindows}
\end{table}

The raw \pai and \e meson yields were corrected for secondary \pai mesons, reconstruction efficiency, 
and acceptance, to obtain the invariant differential yield \cite{Abelev:2012cn,Abelev:2014ypa,Acharya:2017hyu}. 
The secondary \pai mesons from weak decays or hadronic interactions in the ALICE detector were subtracted by 
estimating the contribution in a cocktail simulation, using measured spectra of relevant particles as input. 
The K$^{0}_{S}$ meson is the largest source of secondary \pai mesons, followed by hadronic interactions.
 The contamination from secondaries is largest for low \pT and then steeply decreases with \pT.
This correction is of the order of 8.5\%, 4.4\%, 2.8\%, 7\% at the corresponding lowest \pT and 1.4\%, 2.4\%, $<$~1\%, $<$1~\% at high \pT, 
for PHOS, EMC, PCM-EMC and PCM, respectively, and negligible for PCM$-\gamma^{*}\gamma$.
The PCM analysis is affected by events from bunch crossings other than the triggered one, 
referred to as out-of-bunch pile-up. In the PCM analysis a correction was applied, as described in 
Ref. \cite{Abelev:2014ypa}, that is of the order of 10\% for the lowest \pT bin and sharply declines to about 2\% for high \pT. 
The out-of-bunch pile-up contribution in PHOS, EMC and PCM-EMC is removed by time cuts.
The PCM$-\gamma^{*}\gamma$ analysis used Monte Carlo simulations to apply an additional correction for the remaining 
contamination ($\sim$2.5\%) of the \pai$\rightarrow \gamma\gamma$ in the 
\pai$\rightarrow \gamma^{*}\gamma$ decay channel.
Furthermore, raw \pai and \e meson yield were corrected for acceptance and reconstruction efficiency using 
GEANT3 simulations with HIJING \cite{Gyulassy:1994ew} (PCM and PCM$-\gamma^{*}\gamma$) or 
DPMJET \cite{Roesler:2000he} (PHOS, EMC, PCM and PCM-EMC) as Monte Carlo event generators. 
The product of acceptance and efficiency was calculated in each \pT\ bin and normalized to unit rapidity and full azimuth angle $\Delta\varphi=2\pi$.
A typical value of the acceptance $\times$ efficiency varies from a few percent to few tens of percent, depending on \pT\ and on the reconstruction method.

\subsection{Systematic uncertainties}
\label{sec:sys}

The systematic uncertainties of the \pai and \e invariant differential yields were evaluated as a function of 
 \pT by repeating the analysis for variations on the selection criteria. The magnitude of the systematic uncertainty for 
each set of variations is quantified by the average of the largest significant positive and negative deviations, and is parametrized by a 
low order polynomial function to remove the statistical fluctuations.
\hyperref[tab:SysErrCombPi0]{Table~\ref*{tab:SysErrCombPi0}} and 
\hyperref[tab:SysErrCombEta]{Table~\ref*{tab:SysErrCombEta}} show all the sources of systematic uncertainties and their magnitude 
in two representative \pT bins for \pai and \e mesons, respectively.
All contributions to the total systematic uncertainties within a given 
reconstruction method are considered to be independent and were added in quadrature.  
The systematic uncertainties of the \e/\pai ratio were evaluated independently such that
correlated uncertainties cancel out. All the sources to the total systematic uncertainty are briefly discussed in the following.

For each reconstruction method the material budget is a major source of systematic uncertainty. For the calorimeters 
the uncertainty comes from material in front of the PHOS and EMCal, resulting in $3.5\%$ for PHOS and $4.2\%$ for EMC. 
For the other methods, the material budget 
reflects the uncertainty in the conversion probability of photons \cite{Abelev:2014ffa}, 
adding $4.5\%$ uncertainty for a reconstructed conversion photon.

The yield extraction uncertainty is due to the choice of integration window of the invariant mass distributions. 
The integration window is varied to smaller and larger widths to estimate the error. The yield extraction 
uncertainty for the \pai meson for the different methods is $\sim$2\%, while for the \e meson it increases to $\sim$5\%. 
The yield extraction uncertainty for PHOS is estimated by using the Crystal Ball function instead of a Gaussian to extract the yields, 
resulting in a contribution to the total systematic uncertainty of $2.2\%$ for low \pT and $2.5\%$ for higher \pT.

The PCM $\gamma$ reconstruction uncertainty is estimated by varying the photon quality and 
Armenteros-Podolanski selection criteria. For PCM it is $0.9\%$ at low \pT and increases to $3\%$ for high \pT .
The uncertainty on the identification of conversion daughters in PCM is done by varying the TPC PID 
selection criteria. For PCM it is $0.8\%$ at low \pT and increases to $2.4\%$ for high \pT, 
and for PCM$-\gamma^{*} \gamma$ it is $2.7\%$ at low \pT and decreases to $2.3\%$ for high \pT.
The track reconstruction uncertainty is estimated by varying the TPC track selection criteria. 
This uncertainty slightly increases with increasing \pT and is $\sim$1\%.
The secondary \textit{e$^+$/e$^-$} rejection uncertainty reflects the uncertainty of the real conversion 
rejection from the $\gamma^{*}$ sample and is only present in PCM$-\gamma^{*} \gamma$. 
It is obtained varying the selection on $\psi_{\rm pair}$-$\Delta \varphi$ 
or requiring a hit in the second ITS pixel layer. 
The Dalitz branching ratio uncertainty ($3.0\%$) is taken from the PDG \cite{Olive:2016xmw}.

{\small
\begin{table}[htb]
  \small
  \vspace{0.5cm}
  \centering
    \begin{tabular}{p{3cm}|ll|ll|ll|ll|ll|}
    \cline{2-11}
    &\multicolumn{10}{c|}{\textbf{Relative systematic uncertainty (\%)}}\\
    \cline{2-11}
	 &\multicolumn{2}{c|}{\textbf{PHOS}}&\multicolumn{2}{c|}{\textbf{EMC}}&\multicolumn{2}{c|}{\textbf{PCM}}&\multicolumn{2}{c|}{\textbf{PCM$-\gamma^{*} \gamma$}}&\multicolumn{2}{c|}{\textbf{PCM-EMC}}\\
	 &\multicolumn{2}{c|}{{\pT(GeV/$c$)}}&\multicolumn{2}{c|}{{\pT(GeV/$c$)}}&\multicolumn{2}{c|}{{\pT(GeV/$c$)}}&\multicolumn{2}{c|}{{\pT(GeV/$c$)}}&\multicolumn{2}{c|}{{\pT(GeV/$c$)}}\\
	\cline{2-11}	
	& 1.5 & 7.5 & 1.5 & 14.0 & 1.5 & 7.5 & 0.9 & 3.1 & 1.5 & 7.5\\ \hline
	\multicolumn{1}{|p{3.1cm}|}{Material budget}		& 3.5 & 3.5		& 4.2 & 4.2 	& 9 & 9 	& 4.5 & 4.5 & 5.3 & 5.3\\ 
	\multicolumn{1}{|p{3.1cm}|}{Yield extraction}		& 2.2 & 2.5		& 1.5 & 3.6 	& 2.2 & 1.5 	& 3.5 & 1.1 & 1.2 & 2.6\\ 
	\multicolumn{1}{|p{3.1cm}|}{$\gamma$ reconstruction} 	& &      		& & 		& 0.9 & 3.0 	& 2.3 & 1.8 & 0.6 & 1.7\\ 
	\multicolumn{1}{|p{3.1cm}|}{\textit{e$^+$/e$^-$} identification} & & 		& & 		& 0.8 & 2.4 	& 2.7 & 2.3 & 0.5 & 0.8\\ 
	\multicolumn{1}{|p{3.1cm}|}{Track reconstruction}		& & 		& & 		& 0.3 & 0.7 	& 1.6 & 2.0 & 0.5 & 0.7\\ \hline
	\multicolumn{1}{|p{3.1cm}|}{Sec. \textit{e$^+$/e$^-$} rejection}       & &      & &             &     &         & 4.5 & 2.8 & & \\ 
	\multicolumn{1}{|p{3.1cm}|}{Dalitz branching ratio}               & &           & &             &     &          & 3.0 & 3.0 & & \\ \hline 
	\multicolumn{1}{|p{3.1cm}|}{Cluster energy calib.} 	& 4.9 & 6.2 		& 1.7 & 2.5 	& & 		& & & 2.0 & 2.6\\ 
	\multicolumn{1}{|p{3.1cm}|}{Cluster selection} 		& & 			& 4.6 & 5.1 	& & 		& & & 1.1 & 1.7\\ \hline
	\multicolumn{1}{|p{3.1cm}|}{$\pi^{0}$ reconstruction} 	&  & 			& 0.9 & 3.9 	& 0.9 & 1.1 	&1.9  & 2.0 & 0.3 & 0.3 \\
	\multicolumn{1}{|p{3.1cm}|}{Secondary correction} 	& 1.0  & 			& & & & & & & & \\ 
	\multicolumn{1}{|p{3.1cm}|}{Generator efficiency} 			&  & 		& 2.0 & 2.0 	&  & 	 	&  & 	 & 2.0 & 2.0\\ 
	\multicolumn{1}{|p{3.1cm}|}{Acceptance} 			& 2.2 & 2.2		& & 	&  & 	 	&  & 	 & & \\ 
	\multicolumn{1}{|p{3.1cm}|}{Bkg. estimation} 		& 4.6 & 4.9		& & 		& 0.1 & 0.1 	& 1.8 & 2.0	 &  & \\ 
	\multicolumn{1}{|p{3.1cm}|}{Pile-up correction} 		& 1.0 & 1.0		&  & 	 	& 0.8 & 0.3 	&  &  & & \\ \hline\hline
	\multicolumn{1}{|p{3.1cm}|}{Total} 			& 8.3 & 9.3		& 7.0 & 9.1	& 9.4&	 10.0 	& 9.2 & 7.7 & 6.3 & 7.2\\ \hline

		 	\end{tabular}
			\caption{Relative systematic uncertainties $(\%)$ of the \pai spectrum for the different reconstruction methods. }
			\label{tab:SysErrCombPi0}
\end{table}
}

The uncertainty on the cluster energy calibration is estimated from the relative difference between data and simulation of 
the \pai mass peak position and also includes the uncertainty from the
cluster energy corrections for both calorimeters. 
In the PHOS analysis, the energy calibration is also verified by the
energy-to-momentum $E/p$ ratio of electron tracks reconstructed in the
central tracking system. The residual deviation of \pai mass and $E/p$
ratio of electrons is attributed to the systematic uncertainty of the energy
calibration which contributes $4.9\%$ at low \pT and increases to
$6.2\%$ for high \pT. The uncertainty of the neutral meson spectra caused by the energy calibration uncertainty in
EMC is estimated as $1.7\%$ 
at low \pT and increases to $2.5\%$ for high \pT. The uncertainty on the cluster selection was estimated by varying the minimum energy, minimum number 
of cells and time of the EMCal clusterization process. For the EMC the $\sigma_{\rm long}$ selection and track matching criteria 
are varied to estimate the contribution to the cluster selection uncertainty. This uncertainty accounts for $4.6\%$ at 
low \pT and increases to $5.1\%$ for higher \pT.

The $\pi^{0}$(\e) reconstruction uncertainty is due to the meson selection criteria and was estimated by varying the rapidity 
window of the meson and the opening angle between the two photons. It is a minor contribution to the total error with a 
magnitude of $\sim$1\%. A \pT dependent uncertainty from 2\% at \unit[1]{GeV/$c$} to smaller than 0.5\% for \pT larger than \unit[2]{GeV/$c$} 
is assigned for PHOS to the secondary $\pi^{0}$ correction, and the other methods were not significantly affected by this contribution.
The generator efficiency uncertainty quantifies the difference between different Monte Carlo generators that are used 
to calculate the reconstruction efficiency of the \pai and \e meson and affects photon reconstruction with the EMCal. 
It contributes $2.0\%$ to the \pai meson systematic uncertainty and $4.0\%$ to the \e meson systematic uncertainty.
 The uncertainty on the acceptance correction for PHOS is estimated to be $2.2\%$ and includes the uncertainty introduced by the 
bad channel map. For EMC this uncertainty is included in the generator efficiency correction.

For PCM and PCM$-\gamma^{*} \gamma$, the uncertainty on the background estimation is evaluated by changing the event mixing 
criteria of the photons from using the V$^{0}$ multiplicity to using the charged track multiplicity. For PCM this 
contributes $0.1\%$ ($0.3\%$) for the \pai(\e) meson and for PCM$-\gamma^{*}\gamma$ it contributes $1.8\%$ at 
low \pT and increases to $2.0\%$ for high \pT. For PHOS, the uncertainty of the background is estimated using different polynomial functions
to scale the mixed event background. The contribution is of the order of $4.6\%$, increasing slightly towards high \pT.
The systematic uncertainty due to the out-of-bunch pile-up subtraction is $1.0\%$ for PHOS and it varies from $3.0\%$ at \unit[0.35]{GeV/$c$} 
to $0.3\%$ at high \pT for PCM.

{\small
\begin{table}[hbt]
  \small
  \vspace{0.5cm}
  \centering
    \begin{tabular}{p{3cm}|ll|ll|ll|}
    \cline{2-7}
    &\multicolumn{6}{c|}{\textbf{Relative systematic uncertainty (\%)}}\\
    \cline{2-7}
	 &\multicolumn{2}{c|}{\textbf{EMC}}&\multicolumn{2}{c|}{\textbf{PCM}}&\multicolumn{2}{c|}{\textbf{PCM-EMC}}\\
	 &\multicolumn{2}{c|}{{\pT(GeV/$c$)}}&\multicolumn{2}{c|}{{\pT(GeV/$c$)}}&\multicolumn{2}{c|}{{\pT(GeV/$c$)}}\\
	\cline{2-7}	
	& 3.75 & 14.0 & 1.6 & 7.0 & 3.75 & 7.0 \\ \hline
	\multicolumn{1}{|p{3.1cm}|}{Material budget}		& 4.2 & 4.2 			& 9 & 9 	& 5.3 & 5.3\\ 
	\multicolumn{1}{|p{3.1cm}|}{Yield extraction}		& 6.6 & 8.5 			& 3.1 & 6.1 	& 4.4 & 5.2\\ 
	\multicolumn{1}{|p{3.1cm}|}{$\gamma$ reconstruction} 	& & 				&  3.0 & 5.2 	& 3.0 & 4.2\\ 							
	\multicolumn{1}{|p{3.1cm}|}{\textit{e$^+$/e$^-$} identification} & & 			& 1.8 & 3.4 	& 1.9 & 2.6\\ 
	\multicolumn{1}{|p{3.1cm}|}{Track reconstruction}		& & 			& 1.4 & 1.4		& 2.0 & 2.2\\ \hline
	\multicolumn{1}{|p{3.1cm}|}{Cluster energy calib.} 	& 4.2 & 6.4 			& & 		& 3.5 & 4.7\\ 
	\multicolumn{1}{|p{3.1cm}|}{Cluster selection} 		& 4.9 & 6.7 			& & 		& 3.0 & 3.8\\ \hline
	\multicolumn{1}{|p{3.1cm}|}{$\eta$ reconstruction} 	& 1.6 & 4.1 			& 0.6 &5.6  		& 1.5 & 1.5 \\ 
	\multicolumn{1}{|p{3.1cm}|}{Generator Eff.} 			& 4.0 & 4.0 		&  & 	 	& 2.0 & 2.0\\ 
	\multicolumn{1}{|p{3.1cm}|}{Bkg. estimation} 		& & 				& 0.3 & 0.3 	& & \\ 
	\multicolumn{1}{|p{3.1cm}|}{Pile-up correction} 		&  & 	 		& 0.8 & 0.3 	& & \\ \hline\hline
	\multicolumn{1}{|p{3.1cm}|}{Total} 			& 11.0 & 14.5			& 10.3 & 13.8	  	& 9.6 & 11.3\\ \hline

		 	\end{tabular}
			\caption{Relative systematic uncertainties $(\%)$ of the \e spectrum for the different reconstruction methods. }
			\label{tab:SysErrCombEta}
\end{table}
}

\subsection{pp reference}
\label{sec:ppRef}
In order to quantify cold nuclear matter effects in \pPb\ collisions, we require inclusive
$\pi^{0}$ and $\eta$ distributions in pp collisions at the same collision energy. However,
such distributions are not available at present for pp collisions at \sqrtS~=~5.02~TeV. Therefore, 
the pp reference was calculated by interpolating between the measured spectra at midrapidity 
at \s \cite{Abelev:2014ypa,Acharya:2017hyu}, at \sL \cite{Abelev:2012cn} and at \sLL \cite{Acharya:2017tlv} assuming a power-law 
behavior for the evolution of the cross section in each \pT bin as a function of $\sqrt{s}$ given by 
$\mbox{d}^{2}\sigma(\sqrtS)/\mbox{d}y\mbox{d}\pT \propto \sqrtS^{\alpha(\pT)}$ \cite{Adam:2016dau}, where the fit parameter 
$\alpha(\pT)$ increases with \pT which reflects the hardening of hadron spectra with collision energy.
The method was cross-checked using events simulated by PYTHIA 8.21 
\cite{Sjostrand:2014zea}, where the difference between the 
interpolated and the simulated reference was found to be negligible.

The invariant differential spectra \cite{Acharya:2017hyu,Abelev:2012cn,Acharya:2017tlv}
were fitted either with a Tsallis function~\cite{Tsallis1988,Abelev:2012cn}:
\begin{eqnarray}
\label{eq:tsallis}
  \frac{1}{2\pi N_{\rm ev}}\,\frac{\rm{d^2}
    \it{N}}{\it{p}_{\rm{T}}\rm{d} \it{p}_{\rm{T}}\rm{d}\it{y} }
  =
  \frac{A} {2\pi}\cdot
  \frac{(n-1)(n-2)} {n T(n T+M(n-2))}  \left(1+\frac{m_{T}-M}{nT}\right)^{-n},
\end{eqnarray}
where $M$ is the particle mass, $\mT=\sqrt{M^2 + \pT^2}$, and $A$, $n$
and $T$ are fitting parameters; or with a two component model (TCM) as proposed in Ref.~\cite{Bylinkin:2015xya}: 
\begin{eqnarray}
\frac{1}{2\pi N_{\rm ev}}\,\frac{\rm{d^2}
    \it{N}}{\it{p}_{\rm{T}}\rm{d} \it{p}_{\rm{T}}\rm{d}\it{y} }
= {A}_{\rm {e}} \exp{(-{E}_{\rm {T, kin}}/{T}_{\rm {e}})}+ {A} \left(1 + \frac{\pT^{2}}{{T}^{2} n}   \right)^{-n}
\end{eqnarray}
where ${E}_{\rm {T, kin}}=\sqrt{\pT^2+M^2}-M$ is the transverse kinematic energy of the meson, with $M$ the particle mass,
 ${A}_{\rm {e}}$ and $A$ are normalization factors, ${T}_{\rm {e}}$, $T$ and $n$ are free parameters. 
The parametrizations of the \pai and \e spectra at the different collision energies using the Tsallis or TCM fits were needed due to the different 
\pT binning of the various pp and \pPb spectra. The fits were then evaluated in the used \pPb binning.  The systematic uncertainty for each bin 
was calculated as average uncertainty of adjacent bins in the original binning. 
The statistical uncertainties of the parametrized spectra were computed from the fits to the measured spectra with only statistical errors.

The PHOS, PCM, EMC and PCM-EMC pp references are based solely on their contribution to the published 
spectra \cite{Abelev:2014ypa,Acharya:2017hyu,Abelev:2012cn,Acharya:2017tlv} in order to cancel part of the 
systematic uncertainties in the calculation of \RpPb. The PCM-$\gamma^{*}\gamma$ method used the same pp reference as the PCM. 
The PCM \pai measurement at \s was extrapolated for \pT $>$ 10~\GeVc using the published fit. The PCM  \e measurements  were 
also extrapolated for \pT $>$ 6-8~\GeVc using the published fits.
The difference between the \pai spectrum at $y=0$  and at $y=-0.465$ has been evaluated with PYTHIA 8.21 
to be 1\% for \pT~$>$~\unit[2]{~GeV/$c$} and 0.5\% at \unit[0.5]{~GeV/$c$}.
This correction was applied to the pp reference spectrum. In each \pT bin, the systematic uncertainty of the interpolated 
spectrum was estimated by the largest uncertainty among the input spectra used for the interpolation process.
The statistical error is obtained from the power-law fit.

\section{Results}
\label{sec:results}
\subsection{Invariant yields of \pai and \e mesons}

The ALICE \pai and \e meson invariant differential yields were determined by combining 
the individual meson measurements via a weighted average as described in 
Refs.~\cite{Cowan:1998ji,Nisius:2014wua}. The correlations among the 
measurements for PCM, PCM-EMC, EMC, and PCM$-\gamma^{*}\gamma$ were taken into account using the Best Linear Unbiased 
Estimate (BLUE) method \cite{Lyons:1988rp,Valassi:2003mu}.
The PCM, PHOS and EMC measurements are completely independent and are treated as uncorrelated.
Due to different \pT\ reach, statistics, and acceptance, the binning is not the same for the various methods.
For the combined result, the finest possible binning was chosen. Thus, yields were combined bin by bin 
and methods that did not provide the yield for the specific bin were not taken into account.

The invariant differential meson yields were normalized per NSD event, with the normalization uncertainty 
added in quadrature to the combined systematic uncertainties.
\begin{figure}[phtb]
\begin{center}
    \includegraphics[width=0.45\textwidth]{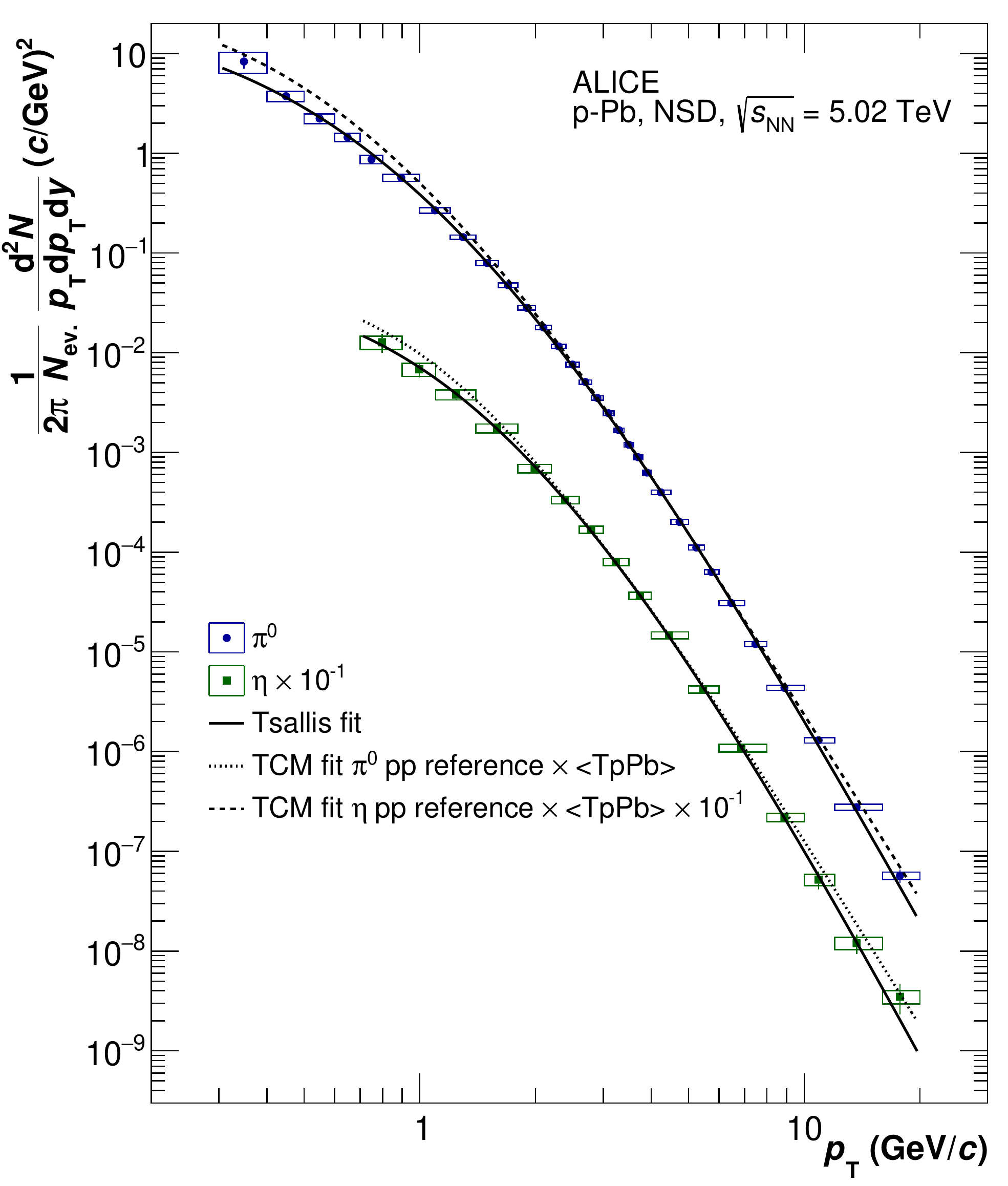}
    \includegraphics[width=0.45\textwidth]{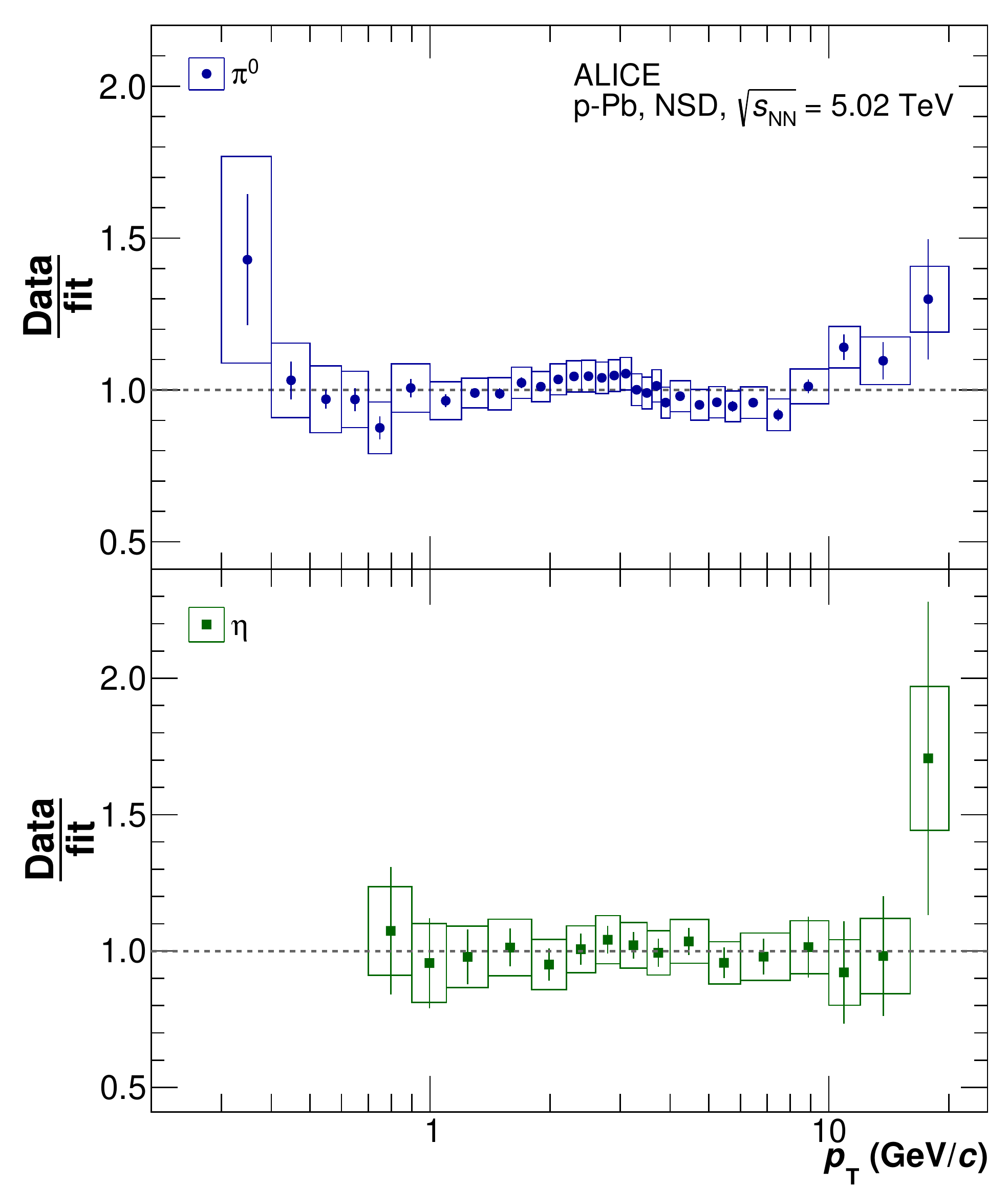}
\end{center}
\caption{Left: Invariant differential \pai and \e yields produced in NSD \pPb collisions at -1.365~$< y_{\mathrm{cms}}<$~0.435 at \spPb. The statistical 
uncertainties are represented as vertical error bars whereas the systematic uncertainties are shown as boxes. 
In addition, Tsallis fits to the measured yields are shown. The TCM fit to the $\langle \TpPb \rangle$ scaled \pai and \e pp 
reference spectra (see \hyperref[sec:RpA]{Sect.~\ref*{sec:RpA}} for details) is shown. Right: Ratios of the measured data to their corresponding Tsallis fits.}
  \label{fig:Pi0EtaYields}
\end{figure}
\begin{table}[pht]
\small
\centering 
\renewcommand{\arraystretch}{1.15}
  \begin{tabular}{p{1.8cm}|c|c|}
  \cline{2-3}
                 & \pai spectrum fit             & \e spectrum fit		\\ \hline
  \multicolumn{1}{|p{1.8cm}|}{  $A$}           & $9.41 \pm 0.49$ & $0.87 \pm 0.10$	\\ 
  \multicolumn{1}{|p{1.8cm}|}{  $n$ }          & $7.168 \pm 0.078$ & $7.56 \pm 0.34$	\\ 
  \multicolumn{1}{|p{1.8cm}|}{  $T$ (GeV/$c$)} & $0.159 \pm 0.004$ & $0.269 \pm 0.019$	\\ \hline 		
  \multicolumn{1}{|p{1.8cm}|}{  $\chi^2/$NDF}  & $0.70$           & $0.18$		\\ \hline	
  \end{tabular}
  \caption{Fit parameters and $\chi^2/$NDF of the Tsallis fits to the combined \pai and combined \e meson invariant differential yields.}
  \label{tab:FitParameters}
\end{table}
\begin{figure}[phtb]
 \includegraphics[width=0.5\textwidth]{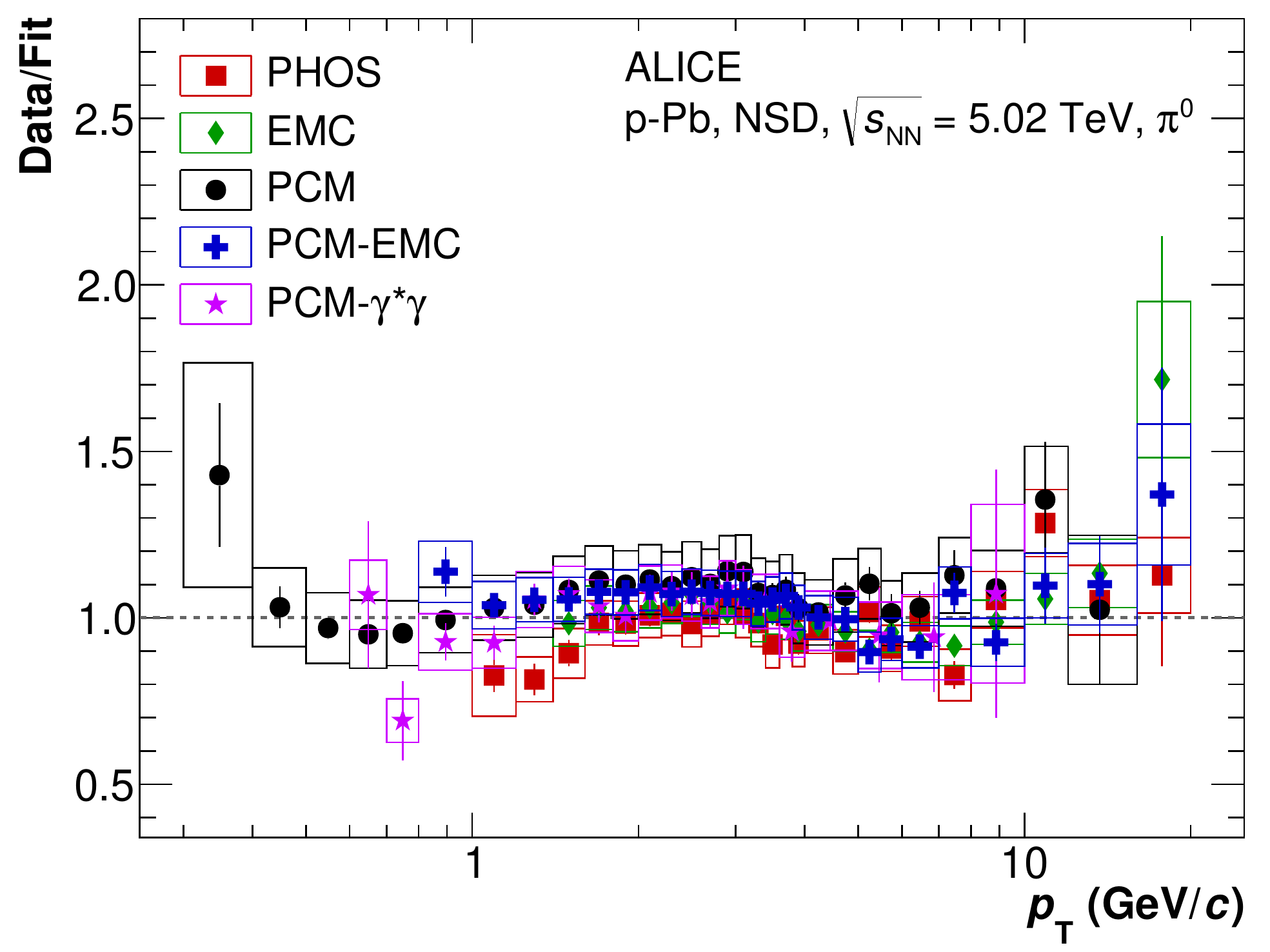}
 \includegraphics[width=0.5\textwidth]{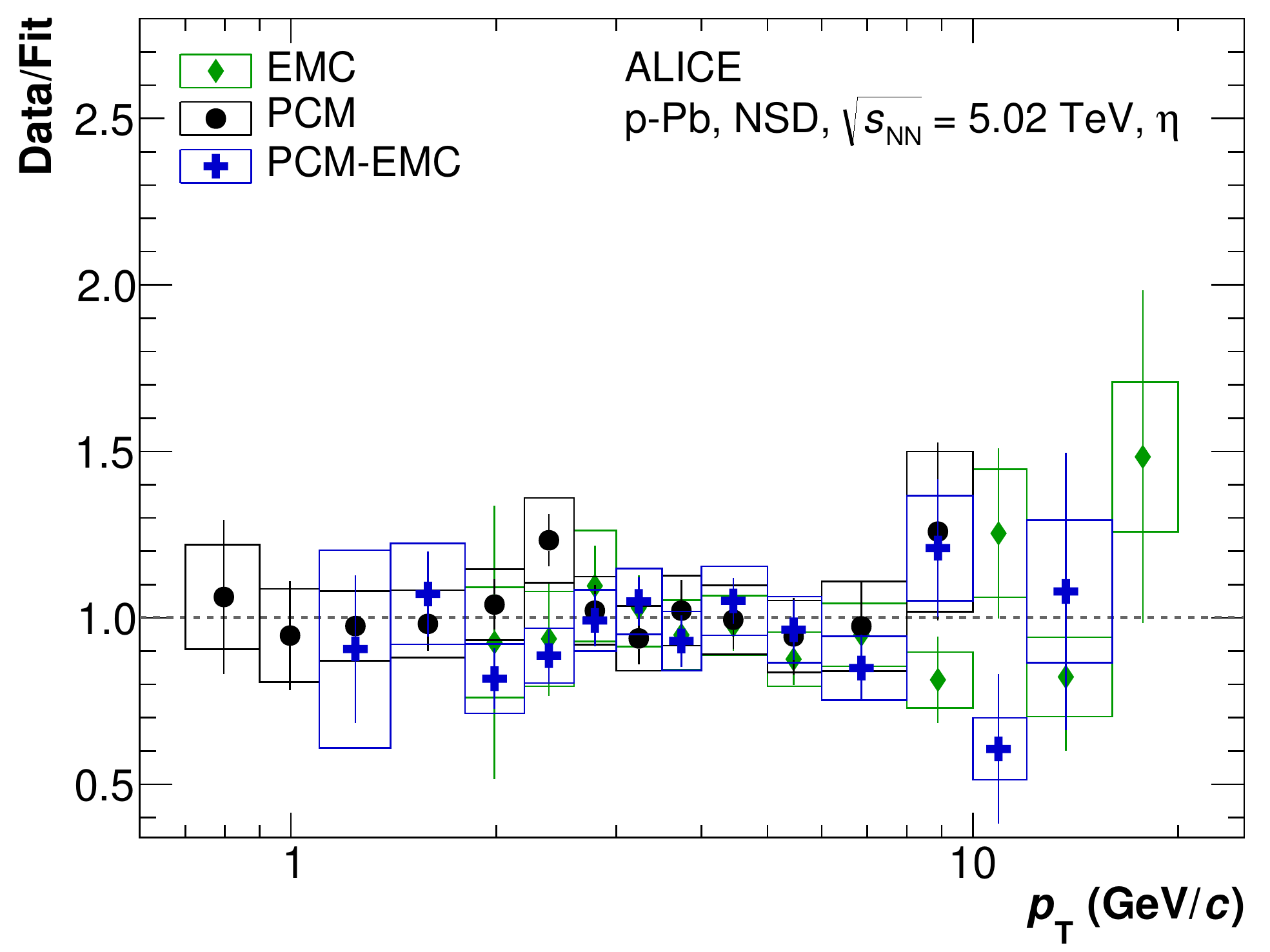}

  \caption{Ratio between individual \pai (left) and \e (right) invariant differential yield measurements, 
and Tsallis fit to the combined meson yield. The statistical uncertainties are represented as vertical 
error bars whereas the systematic uncertainties are shown as boxes.}
  \label{fig:Pi0CombTsallis}		
\end{figure}
The invariant differential \pai and \e yields measured in NSD \pPb collisions at \spPb are shown 
in \hyperref[fig:Pi0EtaYields]{Fig.~\ref*{fig:Pi0EtaYields}}. 
The horizontal location of the data points is shifted towards lower \pT
from the bin center by a few MeV and illustrates the \pT value
where the differential cross section is equal to the measured 
integral of the cross section over the corresponding bin \cite{LAFFERTY1995541}.
For the \e/\pai ratio and \RpPb the bin-shift correction is done in y-coordinates.
Fits with a Tsallis function \hyperref[eq:tsallis]{(Eq. \ref*{eq:tsallis})}
to the combined NSD \pai and \e spectra with statistical and
systematic uncertainties added in quadrature are also shown in 
\hyperref[fig:Pi0EtaYieldsModels]{Fig.~\ref*{fig:Pi0EtaYields}}. 
In each case the Tsallis fit leads to a good description of the meson yield.
The resulting fit parameters and the $\chi^2/$NDF are listed in \hyperref[tab:FitParameters]{Table~\ref*{tab:FitParameters}} 
for the \pai and \e meson. The small values of $\chi^2/$NDF arise from the correlation of systematic uncertainties.
The ratios between the meson yields obtained in the various reconstruction methods and the Tsallis fit to the 
combined spectrum for \pai and \e 
are presented in \hyperref[fig:Pi0CombTsallis]{Fig.~\ref*{fig:Pi0CombTsallis}}. All 
measurements are consistent within uncertainties over the entire \pT range. The invariant
differential yield of neutral pions is consistent with that of charged
pions \cite{Adam:2016dau} over the entire \pT range.

\subsection{ \e/\pai ratio and \mT scaling}

A combined \e/\pai ratio was calculated and is presented in 
\hyperref[fig:EtaPi0RatioALICE]{Fig.~\ref*{fig:EtaPi0RatioALICE}}. 
For this purpose, the \pai was measured with the same binning as the \e meson with the PCM, EMC and PCM-EMC methods. 
The \e/\pai ratio was determined for each method separately to cancel out the common systematic uncertainties 
and then combined taking into account the correlations among the measurements using the BLUE method.
The \e/\pai ratio increases with \pT and reaches a plateau of 0.483~$\pm$~0.015$_{\rm stat}$~$\pm$~0.015$_{\rm sys}$ for \pT $>$ \unit[4]{GeV/$c$}. 
This value agrees with the \e /\pai ratio of $0.48 \pm 0.03$ ($0.47 \pm 0.03$) for \pT $>$ \unit[2]{GeV/$c$} measured 
by PHENIX \cite{PhysRevC.75.024909} in pp (d-Au) collisions at \sNNR and with results from pA collisions at fixed-target 
experiments E515 \cite{Delchamps:1985mi} (p-Pb at $\sqrt{s}=23.8$~GeV, $\e/\pai=0.47\pm 0.03$) and 
E706 \cite{Apanasevich:2002wt} (p-Be at $\sqrt{s}=31.6$~GeV, $\e/\pai=0.45\pm 0.01$ and at $\sqrt{s}=38.8$~GeV, $\e/\pai=0.42\pm 0.01$).
A comprehensive compilation of all measured \e /\pai ratios \cite{PhysRevC.75.024909} shows 
that this ratio reaches an asymptotic value of $R_{\e /\pai}\sim0.4-0.5$ at high \pT in hadronic collisions.
 \hyperref[fig:EtaPi0RatioALICE]{Figure~\ref*{fig:EtaPi0RatioALICE}} shows a good agreement between the \e /\pai ratio 
 measured in \pPb and pp collisions at \spp and \sL with ALICE \cite{Abelev:2012cn}, respectively.
To illustrate universality of the \e /\pai ratio and its independence of the collision system or energy, 
\hyperref[fig:EtaPi0RatioALICE]{Fig.~\ref*{fig:EtaPi0RatioALICE}} also shows the \e /\pai ratio measured 
in \dAu collisions at \sNNR with PHENIX \cite{PhysRevC.75.024909} and  in fixed-target \pBe and \pAu collisions 
at \sNNSPS by the joint TAPS/CERES collaboration \cite{Agakishiev:1998mw} in their corresponding \pT coverage.

\begin{figure}[h]
\begin{center}
\includegraphics[width=0.49\textwidth]{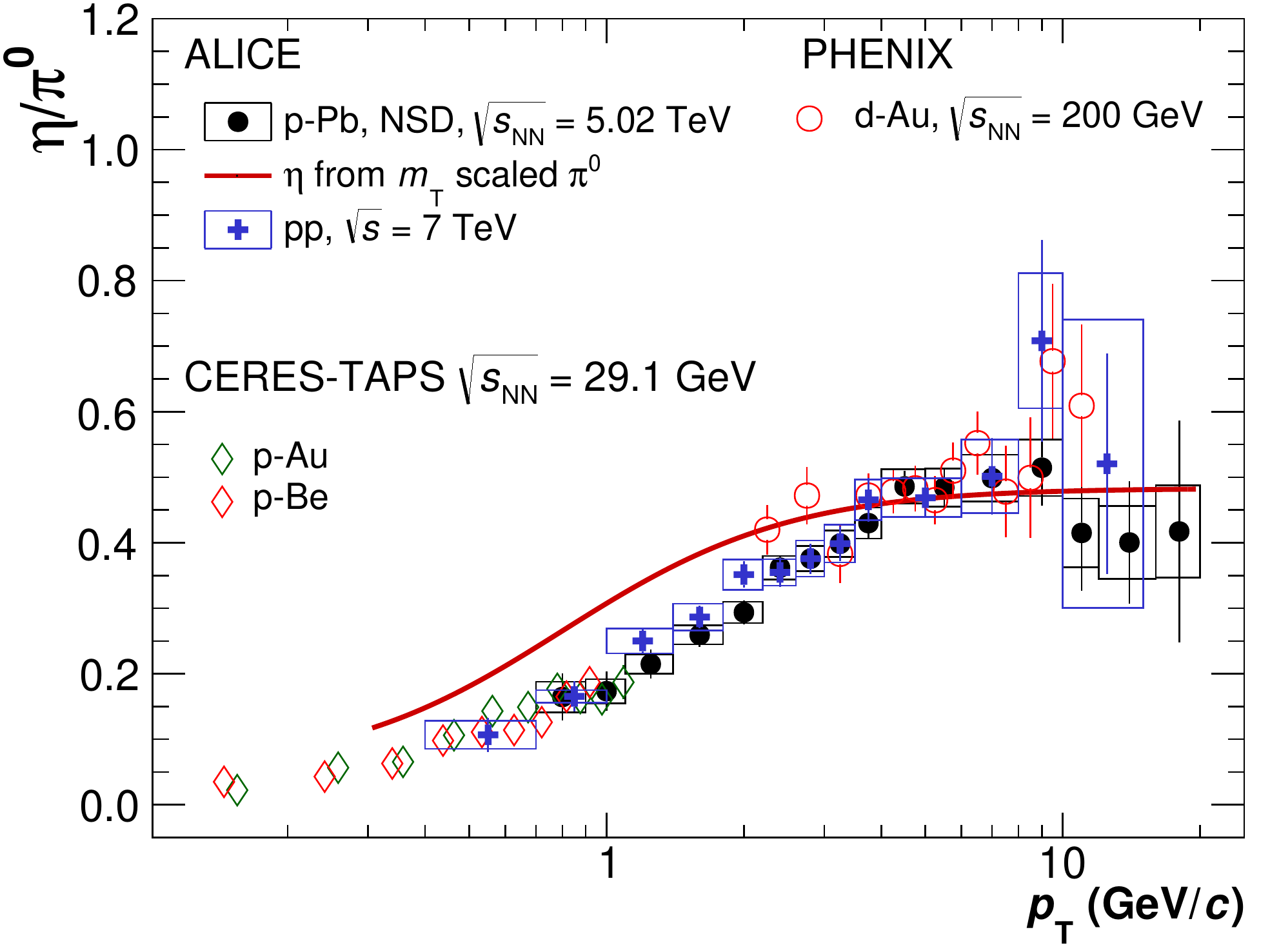}
\includegraphics[width=0.49\textwidth]{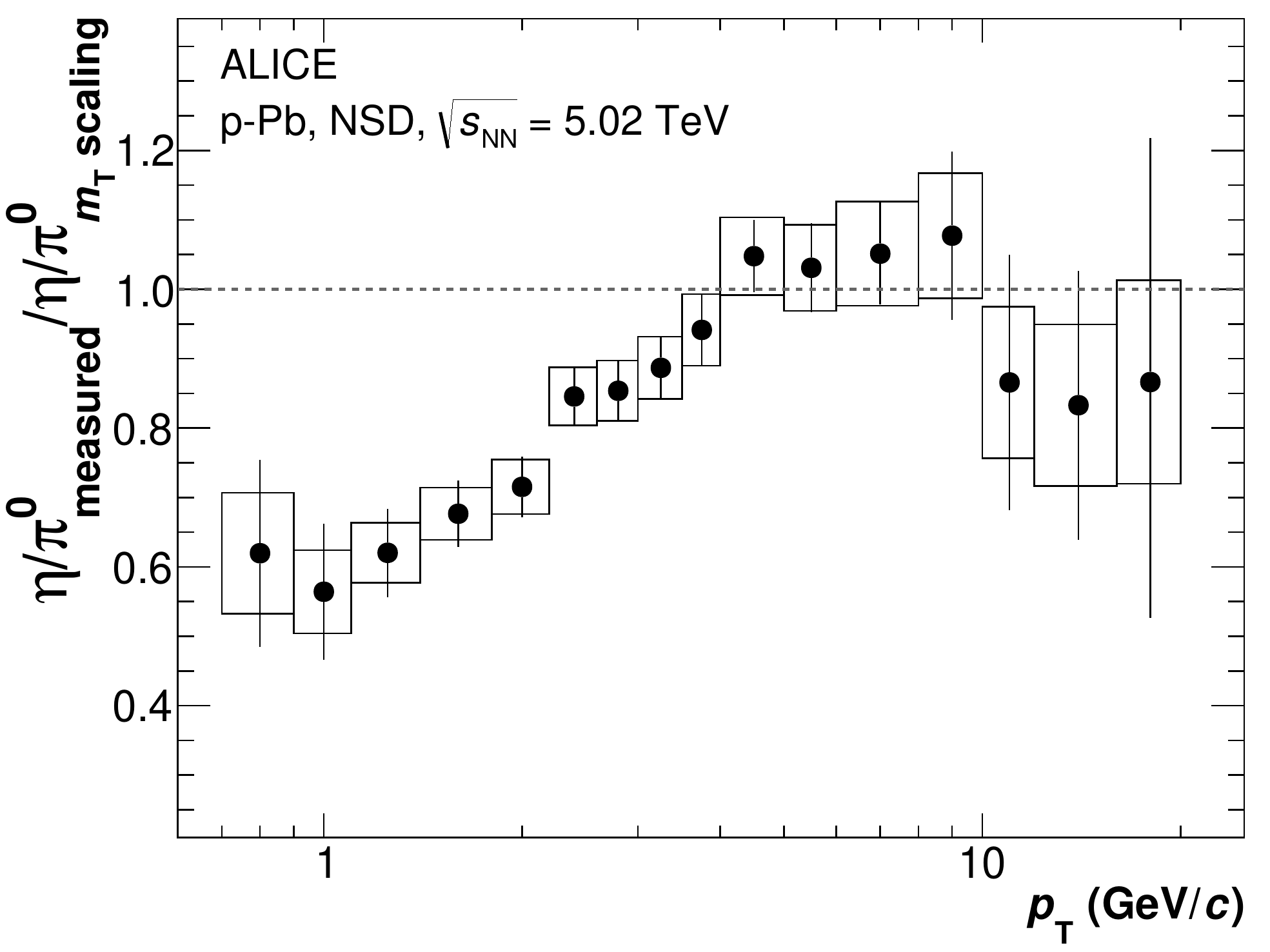}
\end{center}
\caption{Left: \e /\pai ratio as function of \pT measured in NSD \pPb collisions at -1.365~$< y_{\mathrm{cms}}<$~0.435 at \spPb.
The statistical uncertainties are shown as vertical error bars. The systematic uncertainties are represented as boxes. 
For comparison, also the \e /\pai ratios measured in \unit[7]{TeV} pp collisions with ALICE \cite{Abelev:2012cn}, 
in d-Au collisions at \sNNR with PHENIX \cite{PhysRevC.75.024909}, and in \pAu and \pBe collisions at \sNNSPS 
with TAPS/CERES \cite{Agakishiev:1998mw} are shown, as well as the ratio where the \e yield is obtained 
via \mT scaling from the measured \pPb \pai yield. 
Right: Ratio of the measured \e /\pai ratio to the one obtained via \mT scaling.}
  \label{fig:EtaPi0RatioALICE}
\end{figure}

To test the validity of \mT scaling, a comparison of the measured ratio to the ratio 
obtained via \mT scaling is shown in \hyperref[fig:EtaPi0RatioALICE]{Fig.~\ref*{fig:EtaPi0RatioALICE}}.
For this purpose, the \e yield was calculated from the Tsallis parametrization to the combined \pai yield, $P_{\pi^{0}}$,  
assuming \mT scaling $E\,\mbox{d}^3 N^{\eta}/\mbox{d}p^{3} = C_{m}\cdot P_{\pi^{0}}\left(\sqrt{p^2_{\rm{T}}+m^2_{\eta}}\right)$, 
with $C_{m}$ = 0.483~$\pm$~0.015$_{\rm stat}$~$\pm$~0.015$_{\rm sys}$.
The ratio of the \mT-scaled \e yield to the \pai Tsallis fit is 
shown in \hyperref[fig:EtaPi0RatioALICE]{Fig.~\ref*{fig:EtaPi0RatioALICE}} as a red curve.

Above $\pT\sim$ \unit[4]{GeV/$c$} the measured ratio agrees with the \mT-scaled distribution. At 
lower \pT the measured ratio deviates from the \mT scaling prediction, reaching a 40\% difference at \pT~=~1~\GeVc. 
The TAPS/CERES data also supports a deviation from \mT scaling at low \mT while the PHENIX data were found to be consistent 
with \mT scaling, although this measurement starts only at \pT $\sim$ 2 \GeVc. 
The \mT scaling is often 
utilized in measurements of electromagnetic probes \cite{Adam:2015lda,Hayashi:2016vqh} to describe decay photon 
spectra from heavier neutral mesons. 
The measurement reported here demonstrates that \mT scaling is not valid for the \e meson at low \pT.
Therefore, a measured \e yield, especially at low \pT, is crucial for the study of direct photons and dileptons 
in \pA collisions, since \mT scaling from the measured \pai yield overestimates the \e yield at low \pT considerably \cite{Altenkamper:2017qot}.
Measurements of heavier neutral mesons such as $\omega$ in a wide \pT range are thus also desirable.
\\

\subsection{Nuclear modification factor \RpPb}  
\label{sec:RpA}
The ratio of the yield of \pai or \e in \pA collisions relative to that in pp collisions, 
also known as nuclear modification factors (\RpA), are calculated using 
 \begin{equation}
R_{\rm{pPb}} (\pT)=\frac{\mathrm{d}^{2}N^{\text{pPb}}_{\pi^{0},\e}/ \dydpt}{ \langle T_{\text{pPb}} \rangle \cdot \mathrm{d}^{2}\sigma^{\text{pp}}_{\pi^{0},\e}/\dydpt},
\label{eq:RpPb}
\end{equation} 
where $\mathrm{d}^{2}N^{\text{pPb}}_{\pi^{0},\e}/\dydpt$ are the \pai and \e invariant yields measured in \pPb~collisions 
and $d^{2}\sigma^{\text{pp}}_{\pi^{0},\e}/\dydpt$ are the interpolated invariant $\pi^{0}$ and \e meson 
cross sections in pp collisions at \spPb, as described in \hyperref[sec:ppRef]{Sect.~\ref*{sec:ppRef}}. 
$\langle T_{\text{pPb}}\rangle$ 
is the average nuclear overlap function, $\langle \TpPb \rangle = 0.0983\pm 0.0035~\mbox{mb}^{-1}$ 
\cite{PhysRevLett.110.032301,ALICE:2012mj}.

\begin{figure}[hbt]
\begin{center}
\includegraphics[width=0.49\textwidth]{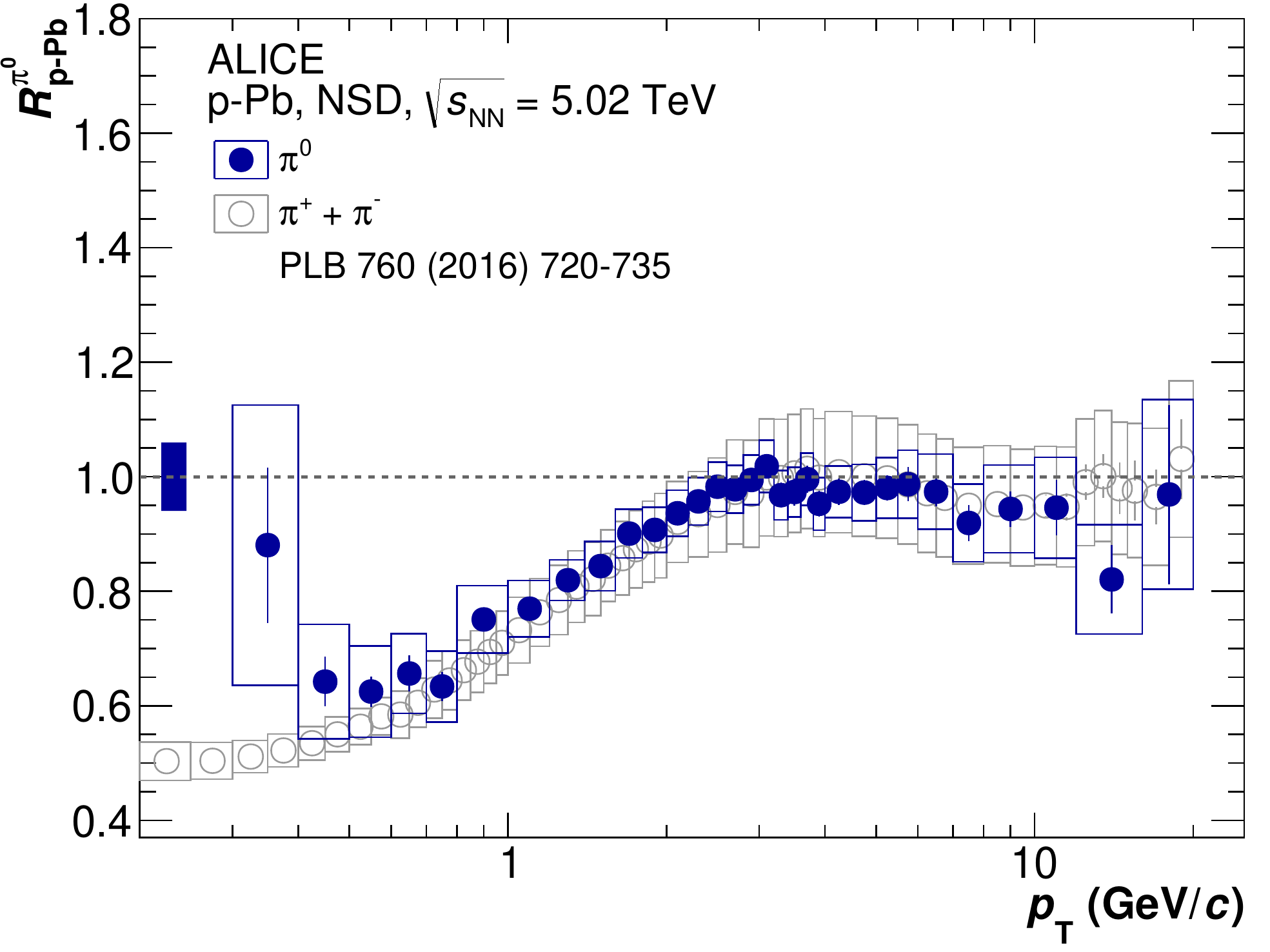} 
\includegraphics[width=0.49\textwidth]{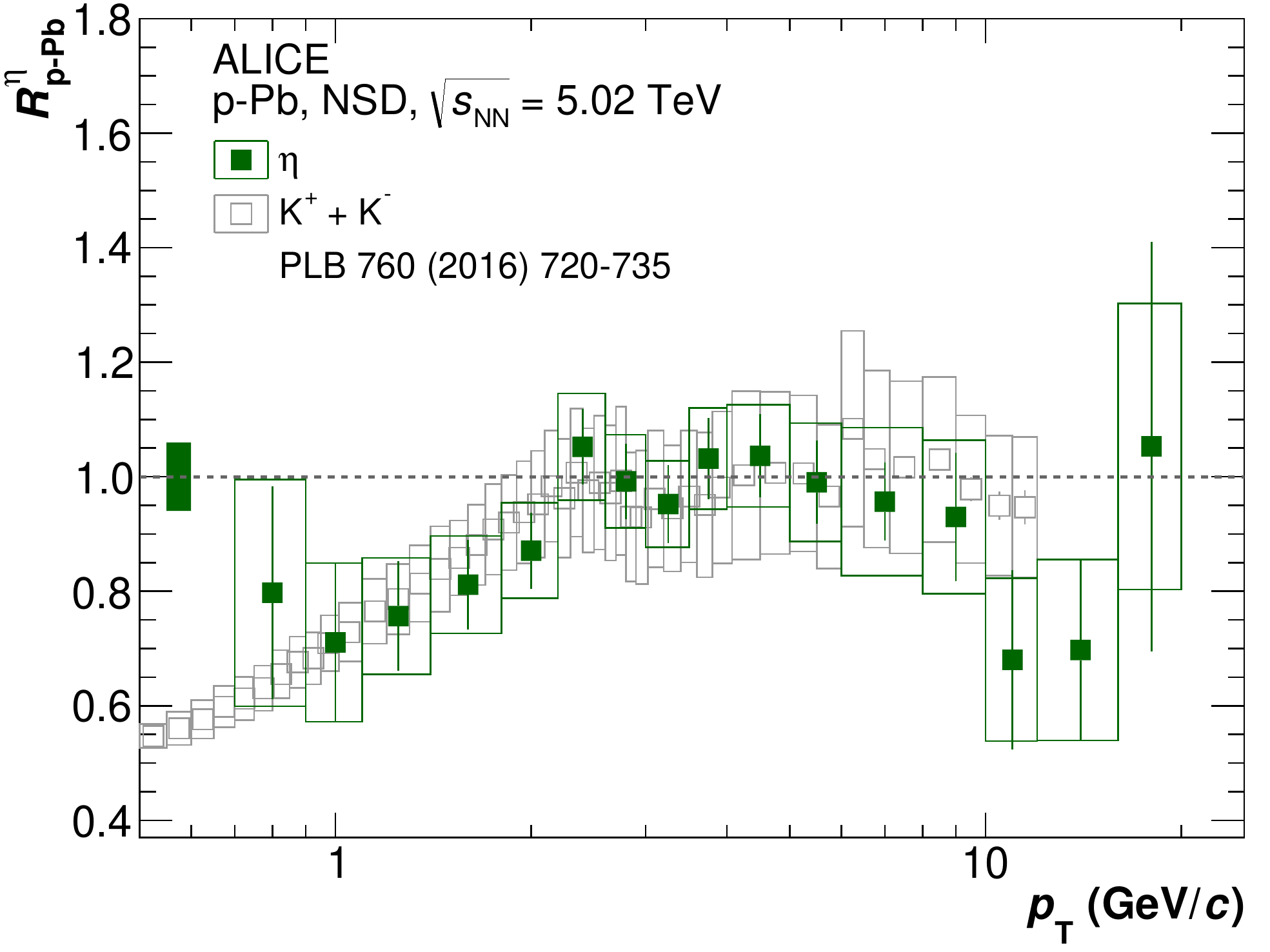} 
\end{center}
\caption{\pai (left) and \e (right) nuclear modification factors \RpPb measured in NSD \pPb collisions 
at -1.365~$< y_{\mathrm{cms}}<$~0.435 at \spPb compared to the nuclear modification factors of charged pions and charged kaons, respectively. 
The statistical uncertainties are shown as vertical error bars and the systematic uncertainties are represented as boxes. 
The overall normalization uncertainty is given as the solid black box 
around unity.}
  \label{fig:RpPb} 
\end{figure}

In the absence of nuclear effects, \RpPb\ is unity in the \pT region 
where hard processes dominate particle production. 
The values of \RpPb\ were calculated for each individual method to cancel out
the common systematic uncertainties and then combined using the BLUE method 
(\hyperref[fig:RpPb]{Fig.~\ref*{fig:RpPb}}). 
For the Dalitz \RpPb\ the PCM pp reference is used. This induces a correlation of the Dalitz \RpPb with the \RpPb from PCM.
The NSD normalization uncertainty is added in quadrature to the
overall normalization uncertainty together with the uncertainties of the \TpPb and of the inelastic pp cross sections. 
The fit to the reference \pai and \e spectra in pp collisions at $\sqrt{s}=5.02$~TeV scaled 
by $\langle \TpPb \rangle$ 
are also displayed in \hyperref[fig:Pi0EtaYields]{Fig.~\ref*{fig:Pi0EtaYields}}. 
The fit parameters are given in \hyperref[tab:FitParametersPPref]{Table~\ref*{tab:FitParametersPPref}}. 

The values of \RpPb are consistent with unity for transverse momenta above \Unit{2}{\GeVc} for the \pai and \e mesons. 
The \RpPb measurements for neutral and charged pions as well as the \RpPb measurements for \e mesons and charged 
kaons \cite{Adam:2016dau} agree with each other within uncertainties over the complete \pT range as shown in (\hyperref[fig:RpPb]{Fig.~\ref*{fig:RpPb}}).

\subsection{Comparisons to theoretical models}


Comparisons of the \pai and \e meson transverse momentum spectra
to several theoretical calculations are shown in \hyperref[fig:Pi0EtaYieldsModels]{Fig.~\ref*{fig:Pi0EtaYieldsModels}}. 
In the following, we discuss each model individually, compared with the experimental data.

\begin{figure}[htb]
\begin{center}
  \includegraphics[width=0.49\textwidth]{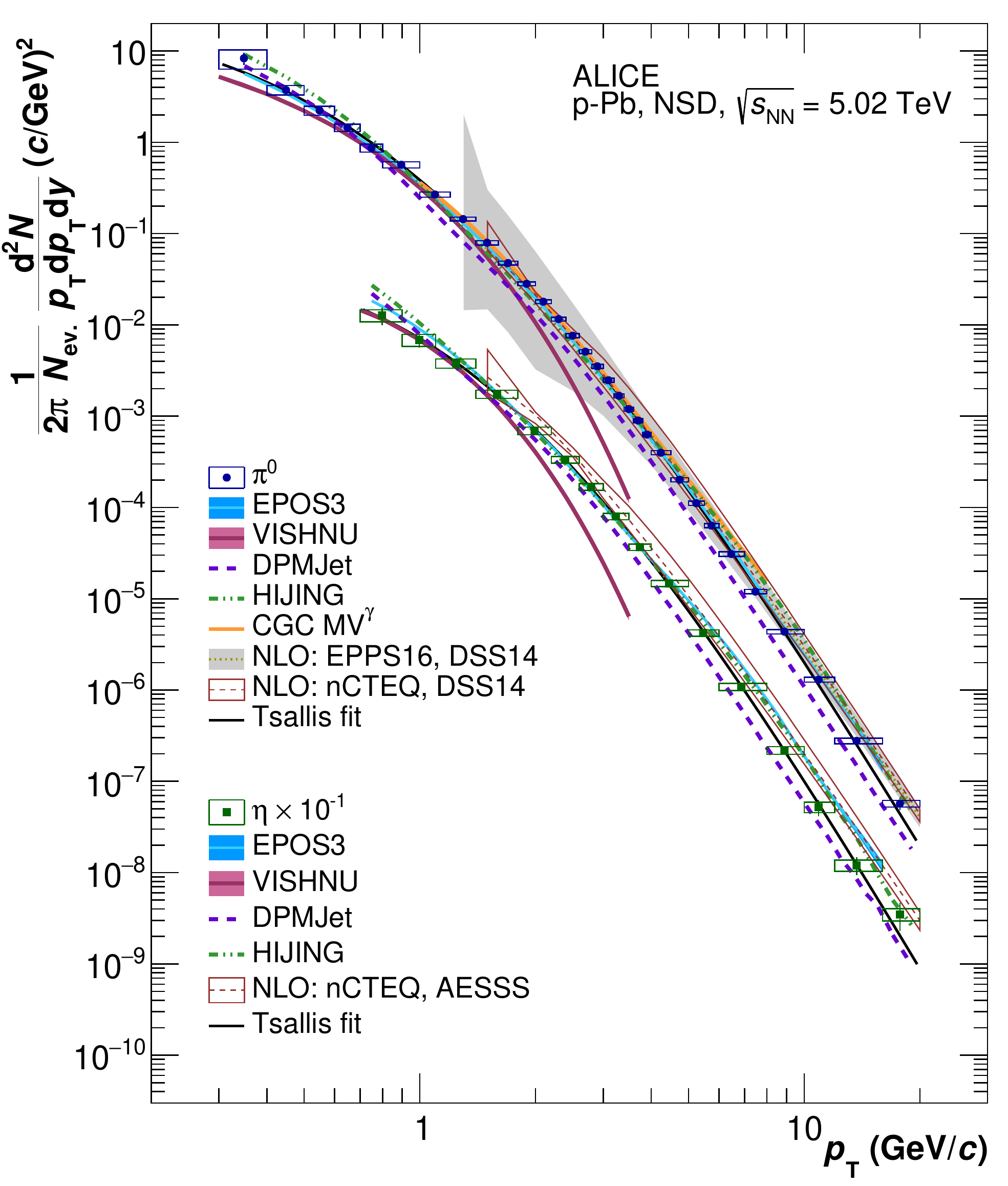}
  \includegraphics[width=0.49\textwidth]{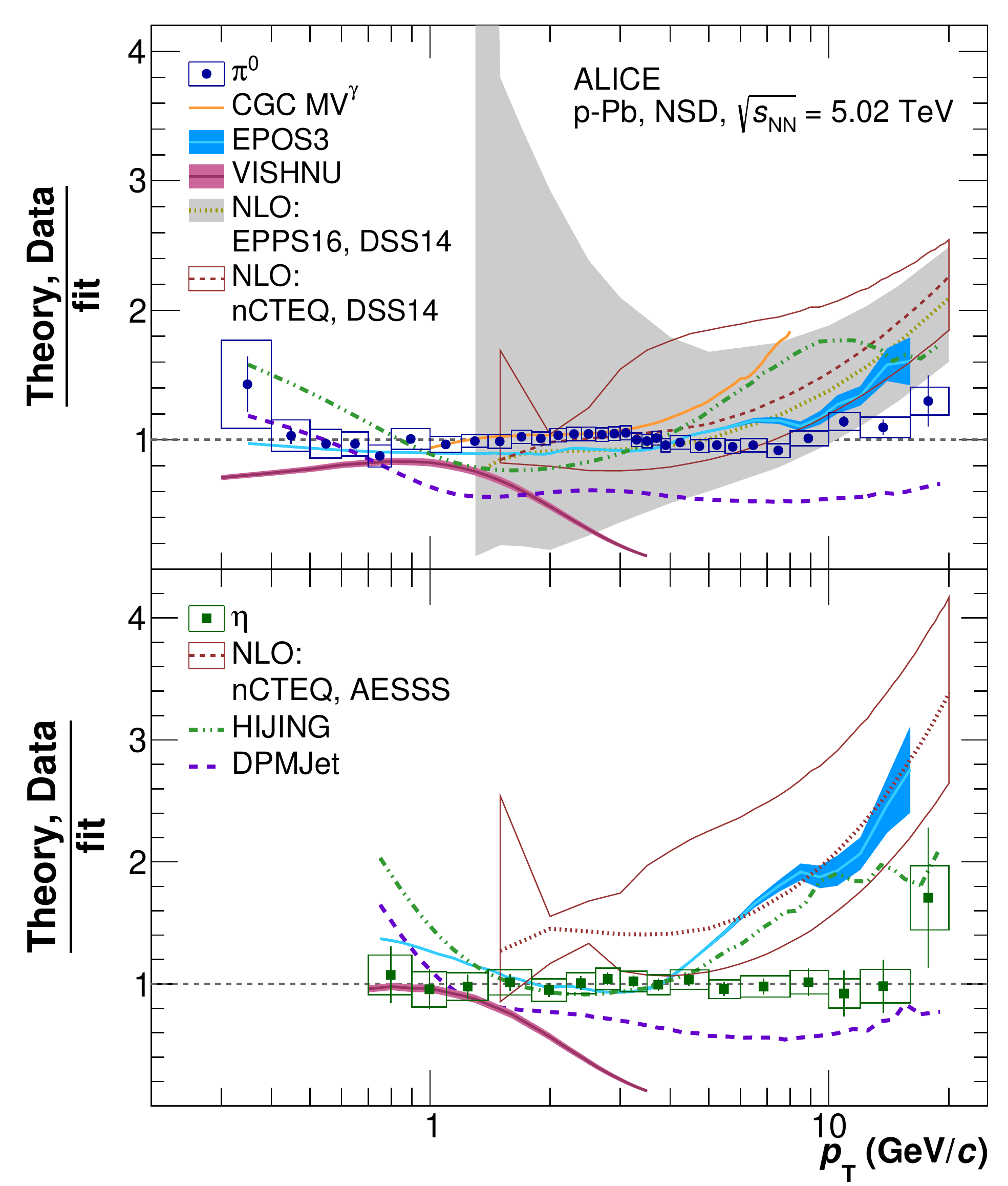}\vspace*{6pt}
\end{center}
\caption{Comparison of several theoretical calculations to the invariant differential \pai and \e yields produced in 
NSD \pPb collisions at -1.365~$< y_{\mathrm{cms}}<$~0.435 at \spPb from \hyperref[fig:Pi0EtaYieldsModels]{Fig.~\ref*{fig:Pi0EtaYields}}. 
Theoretical calculations are shown for the EPOS3 model \cite{Werner:2013tya}, 
CGC model \cite{Lappi:2013zma}, pQCD calculations at NLO \cite{Helenius:2012wd,dEnterria:2013sgr,Jager:2002xm} using 
EPPS16 nPDF \cite{Eskola:2016oht} or using the nCTEQ nPDF \cite{Kovarik:2015cma} and DSS14 FF \cite{deFlorian:2014xna} 
for the \pai and using nCTEQ nPDF \cite{Kovarik:2015cma} and AESSS FF \cite{Aidala:2010bn} for the \e meson, hydrodynamic 
framework (labeled as VISHNU) \cite{Shen:2016zpp} using the iEBE-VISHNU code \cite{Shen:2014vra}, 
DPMJET model \cite{Roesler:2000he}, and HIJING model \cite{Gyulassy:1994ew}. The blue band on the EPOS3 calculation
shows the statistical errors of the prediction. The gray band on the pQCD calculation includes the uncertainties on the 
factorization, renormalization and fragmentation scales, as well as on the nPDF and FF.
 The ratios of the measured data and several theoretical calculations to the data Tsallis 
fits (\hyperref[fig:Pi0EtaYieldsModels]{Fig.~\ref*{fig:Pi0EtaYields}}) are shown in the right panel.}
  \label{fig:Pi0EtaYieldsModels}
\end{figure}

pQCD calculations at NLO \cite{Helenius:2012wd,dEnterria:2013sgr,Jager:2002xm} using the EPPS16 
nPDF \cite{Eskola:2016oht} with the CT14 PDF \cite{Dulat:2015mca} or using the 
nCTEQ nPDF \cite{Kovarik:2015cma} and DSS14 FF \cite{deFlorian:2014xna} reproduce 
the \pai spectrum in \hyperref[fig:Pi0EtaYieldsModels]{Fig.~\ref*{fig:Pi0EtaYieldsModels}}, 
within the uncertainties due to the nPDF, the FF and variation of the factorization, 
renormalization and fragmentation scales. 
The largest contribution to the systematic uncertainty is due to the uncertainties in the choice of scales.
Note that the EPPS16 nPDF has larger uncertainties than EPS09 nPDFs. 
pQCD calculations at NLO \cite{Jager:2002xm} using the nCTEQ nPDF \cite{Kovarik:2015cma} and AESSS FF \cite{Aidala:2010bn} 
reproduce the \e meson spectrum at intermediate \pT while it overestimates the spectrum up to a factor two at high \pT.
Inclusive \e meson production has been measured in pp collisions at different LHC energies 
\cite{Abelev:2012cn,Acharya:2017hyu,Acharya:2017tlv}, which could be used to improve 
the \e meson FF \cite{Aidala:2010bn} utilizing global fits, similar to a recent calculation for pions and 
kaons \cite{deFlorian:2014xna,deFlorian:2017lwf}.

The HIJING model \cite{Gyulassy:1994ew} combines a pQCD-based calculation for multiple jet production 
with low \pT multi-string phenomenology. The model includes multiple minijet production, nuclear shadowing of parton 
distribution functions, and a schematic mechanism of jet interactions in dense matter. The Glauber model for multiple 
collisions is used to calculate \pA and \AACol collisions.
\hyperref[fig:Pi0EtaYieldsModels]{Figure~\ref*{fig:Pi0EtaYieldsModels}} 
shows that the central value of the model calculation for inclusive 
\pai is about 20\% smaller than the measured value at intermediate \pT, between 1 and 4~\GeVc, 
while it agrees with the \e meson production in \pPb collisions. 
At lower and higher \pT values the calculation overestimates the \pai and \e yields by up to 60-80\%.

The DPMJET event generator \cite{Roesler:2000he} based on the Gribov-Glauber approach is an implementation 
of the two-component Dual Parton Model. This model treats the soft and the hard scattering processes in an 
unified way, using Reggeon theory for soft processes and lowest
order pQCD for the hard processes. DPMJET was tuned to reproduce RHIC
measurements of hadron production at low and moderate \pT by introducing a new mechanism of
percolation and chain fusion, though it overestimates inclusive hadron yields at
high \pT at RHIC energies \cite{Bopp:2004xn}. Comparison of the \pai and \e
meson measurements with DPMJET calculations in \hyperref[fig:Pi0EtaYieldsModels]{Fig.~\ref*{fig:Pi0EtaYieldsModels}} 
shows that the model reproduces the distributions for \pT $<$ 1~\GeVc, but
underestimates the yields by 40\% at higher \pT. This suggests that the model parameters
may need to be adjusted for the new energy domain.
Comparison of DPMJET model predictions to particle production measurements in pp collisions at LHC energies  
also shows that the energy dependence of hadron production predicted by the model 
does not agree with data \cite{Bopp:2011fw}.

The \pai invariant differential yield computed with the CGC model \cite{Lappi:2013zma} with 
MV$^\gamma$ \cite{Albacete:2010sy} as the initial condition agrees with the measurements in 
\hyperref[fig:Pi0EtaYieldsModels]{Fig.~\ref*{fig:Pi0EtaYieldsModels}} 
for \pT $<$ 5~\GeVc. 
The deviation seen at high \pT is similar to that observed for inclusive \pai production 
in pp collisions at LHC.

The iEBE-VISHNU package \cite{Shen:2014vra} consists of a 3+1 viscous hydrodynamical 
model coupled to a hadronic cascade model \cite{Shen:2016zpp}. 
Fluctuating initial conditions in the transverse plane are generated using a Monte-Carlo Glauber model.
\hyperref[fig:Pi0EtaYieldsModels]{Figure~\ref*{fig:Pi0EtaYieldsModels}} 
shows that this model reproduces the \pai and \e meson inclusive 
spectra for 0.7 $<$ \pT $<$ 1.5~\GeVc. For lower momenta (\pT $<$ 0.7~\GeVc) the model prediction
is lower than the measured \pai yield by up to a factor of two at 0.35~\GeVc.
For \pT $>$~1.5~\GeVc the model predictions underestimate the \pai and \e meson yields by a factor 5 at 3.5~\GeVc. 
This comparison shows that additional mechanisms not included in the model, in particular jet production, 
are important to describe particle production in \pPb collisions in this region.

The EPOS3 \cite{Werner:2013tya} event generator is based on 3D+1 viscous hydrodynamics, with flux tube initial 
conditions that are generated in the Gribov-Regge multiple scattering framework.
The reaction volume is divided into a core and a corona part. The core is evolved using viscous
hydrodynamics. The corona is composed of hadrons from string decays.
\hyperref[fig:Pi0EtaYieldsModels]{Figure~\ref*{fig:Pi0EtaYieldsModels}} 
shows that this model reproduces the \pai inclusive \pT 
spectrum well over the full measured range. The model also reproduces the charged pion and kaon inclusive 
spectra in \pA collisions \cite{Werner:2013tya}. However, the \e meson spectrum is 
well-reproduced only for \pT $<$ 4~\GeVc, while at \pT~$>$~4~\GeVc the calculations lie above the data, 
with the disagreement reaching a factor of two at 10~\GeVc. Note that the VISHNU theoretical predictions \cite{Shen:2016zpp} 
and EPOS3 are within 10-20\% and 30-40\% for the \pai and \e mesons, respectively, for \pT $<$ 1.5~\GeVc.
The comparisons to VISHNU and EPOS3 shows that a picture incorporating viscous hydrodynamic flow is consistent with measured
particle yields at low \pT in \pPb collisions.

Comparison of the measured high-precision \pai and \e meson spectra with theoretical 
models in \hyperref[fig:Pi0EtaYieldsModels]{Fig.~\ref*{fig:Pi0EtaYieldsModels}} 
clearly shows that different underlying pictures can describe the data qualitatively. However,
systematic uncertainties of the theoretical models are not provided, or are sizable. 
Hydrodynamic models agree with the data at low \pT, while jet production appears to be needed 
for a good description at $\pT>4$~\GeVc. While the high \pT part of the spectra can be described 
by NLO pQCD calculations, the precise data presented here will help to reduce their uncertainties significantly, 
for instance providing additional constraints on identified-particle FFs. 
Improved theoretical uncertainties are needed in order to discriminate among the models.

The comparison of the \e/\pai ratio to different theoretical predictions is shown in
\hyperref[fig:EtaPi0RatioALICEModels]{Fig.~\ref*{fig:EtaPi0RatioALICEModels}}. 
The DPMJET and HIJING model calculations are very close to the \mT scaling prediction, 
i.e. they lie above the measured ratio for \pT $<$ 4~\GeVc, and agree with it at larger \pT. 
On the other hand, the EPOS3 model calculation is closer to the data at low \pT than the \mT scaling prediction, while 
for \pT~$>$~4~\GeVc it continues to increase instead of reaching the plateau observed in data. 
The prediction from the VISHNU hydrodynamical calculation is in agreement with the measured data and very 
close to the EPOS3 prediction. However, this comparison may only be relevant up to \pT of 1.5~\GeVc where the calculation 
was able to reproduce the measured neutral meson spectra.
This behavior highlights once more the importance of hydrodynamical flow in \pPb collisions at the LHC.
\begin{figure}[htb]
\begin{center}
\includegraphics[width=0.65\textwidth]{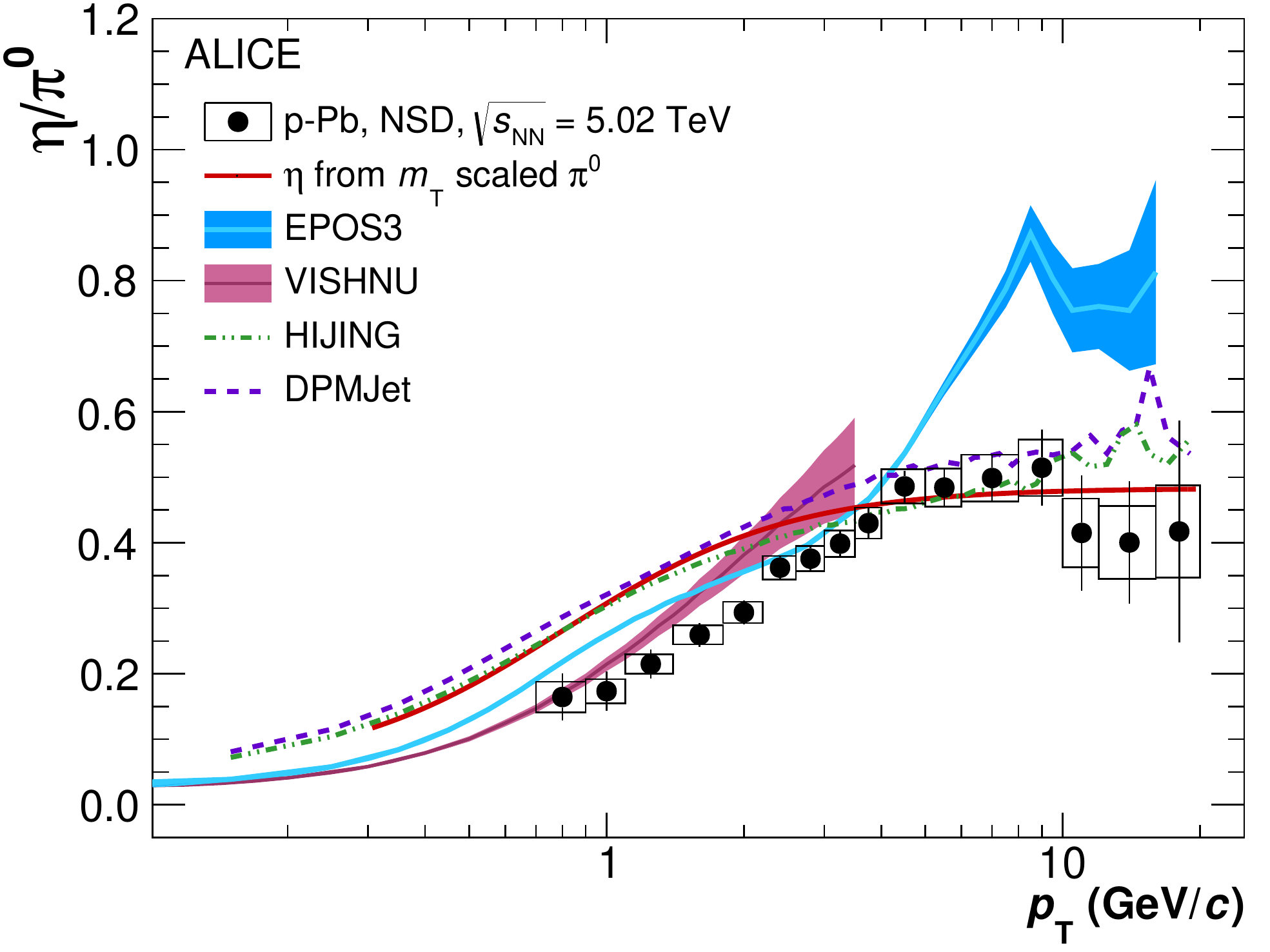}
\end{center}
\caption{Comparison of different theoretical calculations to the \e /\pai ratio measured in NSD \pPb collisions 
at -1.365~$< y_{\mathrm{cms}}<$~0.435 at \spPb from \hyperref[fig:EtaPi0RatioALICE]{Fig.~\ref*{fig:EtaPi0RatioALICE}}.
Theoretical calculations are shown for the EPOS3 
model \cite{Werner:2013tya} with statistical errors shown as a band, hydrodynamic framework (VISHNU) \cite{Shen:2016zpp} 
using the iEBE-VISHNU code  \cite{Shen:2014vra}, DPMJET model \cite{Roesler:2000he} and HIJING model \cite{Gyulassy:1994ew}.}
\label{fig:EtaPi0RatioALICEModels}
\end{figure}

A comparison of the measured \pai and \e \RpPb to different model predictions is shown in 
\hyperref[fig:RpPbModels]{Fig.~\ref*{fig:RpPbModels}}.
\begin{figure}[hbt]
\begin{center}
\includegraphics[width=0.49\textwidth]{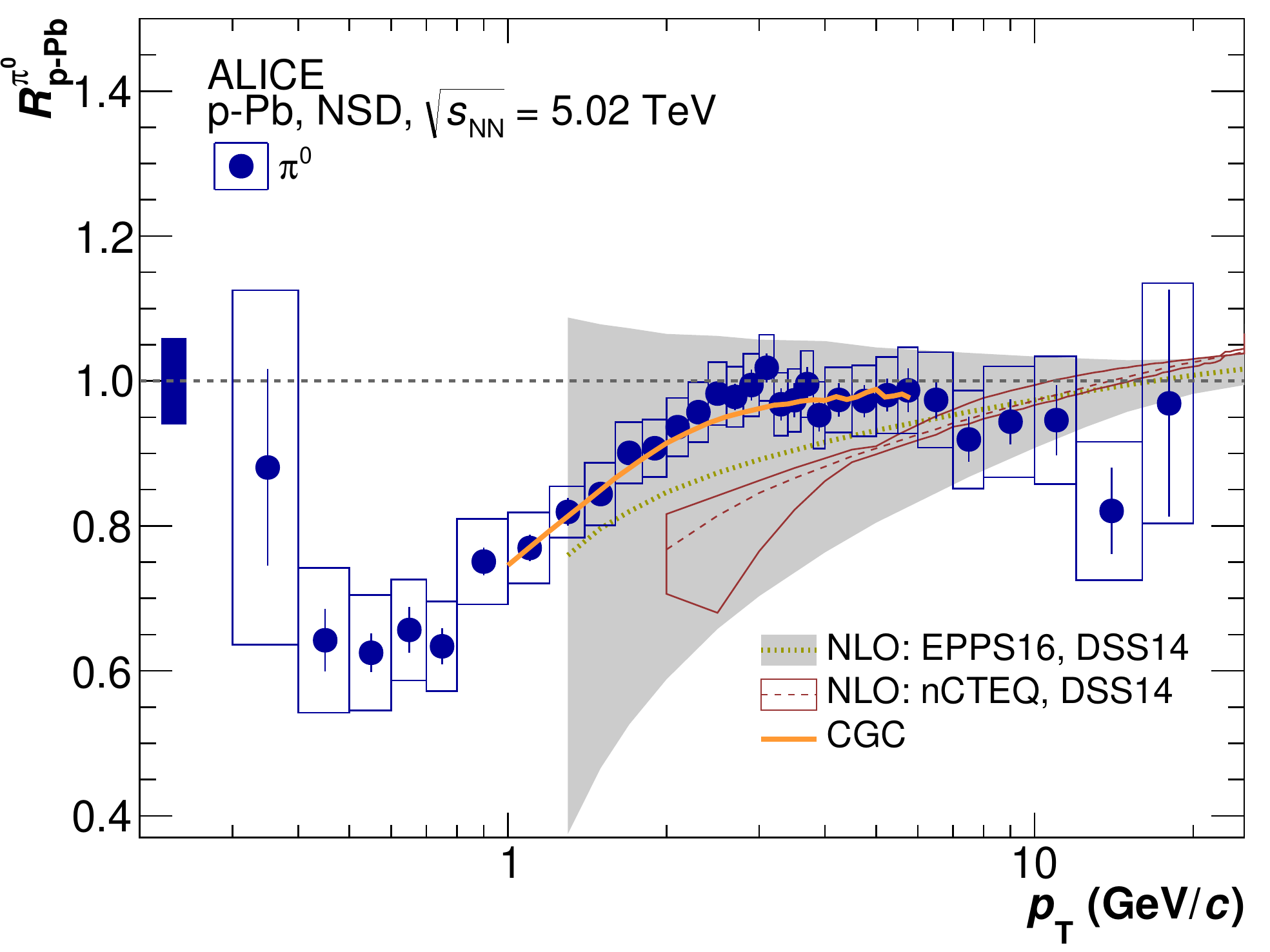} 
\includegraphics[width=0.49\textwidth]{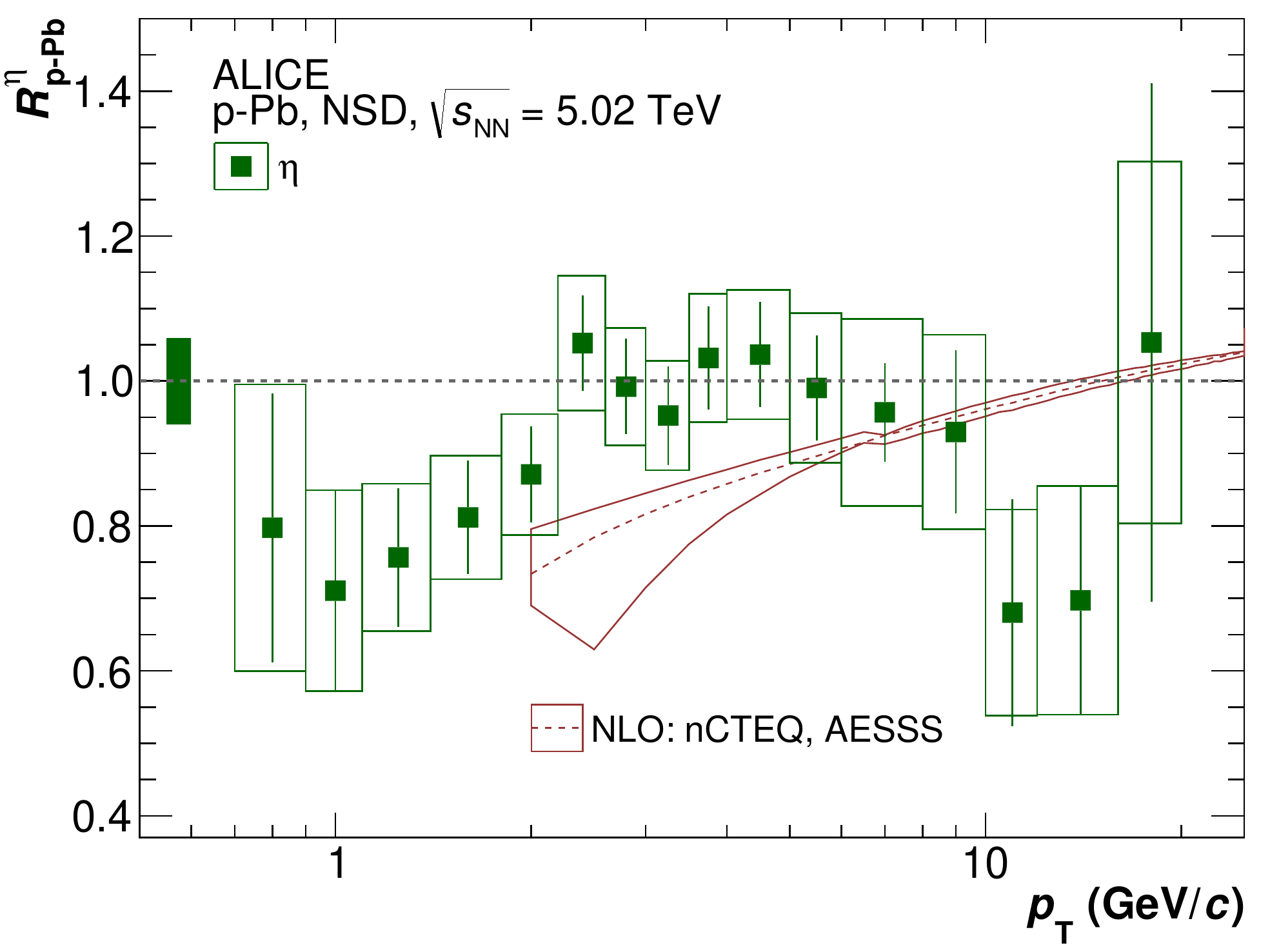} 
\end{center}
\caption{Comparison of different theoretical model calculations to the \pai (left) and \e (right) nuclear 
modification factors \RpA measured in NSD \pPb collisions  at -1.365~$< y_{\mathrm{cms}}<$~0.435 at \spPb. 
The grey band shows a pQCD calculation at NLO using the EPPS16 nPDF~\cite{Eskola:2016oht}, 
the CT14 PDF~\cite{Dulat:2015mca} and the DSS14 FF~\cite{deFlorian:2014xna} including systematic uncertainties. 
Color Glass Condensate predictions using the {\ensuremath{k_{\mbox{\tiny T}}}\xspace} factorization method are also shown. 
NLO calculations using nCTEQ nPDF \cite{Kovarik:2015cma}, and DSS14 FF  (\pai) \cite{deFlorian:2014xna} 
or  AESSS FF (\e) \cite{Aidala:2010bn} are also shown.}
  \label{fig:RpPbModels} 
\end{figure}
 The NLO pQCD calculations for the \pai \cite{Helenius:2012wd,dEnterria:2013sgr,Jager:2002xm}
utilize the EPPS16 nuclear PDF \cite{Eskola:2016oht} or the nCTEQ nPDF \cite{Kovarik:2015cma}, 
and DSS14 FF \cite{deFlorian:2014xna}, and for the \e meson \cite{Jager:2002xm} the nCTEQ 
nPDF \cite{Kovarik:2015cma} and AESSS FF \cite{Aidala:2010bn} are used.
The central values of the NLO predictions for \pai and \e lie below the data for \pT $<$ \unit[6]{GeV/$c$}. 
While the uncertainties of \pai calculations using nCTEQ are small and show sizable difference, the uncertainties 
for \pai calculations using EPPS16 are large and in agreement with the data.
The CGC prediction from Ref. \cite{Lappi:2013zma} uses the 
{\ensuremath{k_{\mbox{\tiny T}}}\xspace} factorization method and is able to reproduce the measured \RpPb.

\section{Conclusions}
\label{sec:conclusions}
The \pT differential invariant yields of \pai and \e mesons were measured in NSD
\pPb collisions at \spPb in the transverse momentum range $0.3 < \pT < 20$~\GeVc and $0.7 < \pT<20$~\GeVc, 
respectively.
State-of-the-art pQCD calculations at NLO are able to describe the \pai spectrum within the 
uncertainties of the nPDF and the pQCD scale, whereas they describe the \e spectrum at intermediate \pT and overestimate it
up to a factor of two at high \pT. As the wealth of the \e measurements is already sizable at the LHC, 
it will be important to include them in global fits to reach a similar theoretical progress in the pQCD calculations 
of the \e meson.
  
The \e/\pai ratio is constant with a value of 0.483~$\pm$~0.015$_{\rm stat}$~$\pm$~0.015$_{\rm sys}$ at $\pT > 4$~\GeVc which is consistent 
with the \e/\pai measurements at lower-energy pp, \pA and AA collisions.
Universality of the \e/\pai behavior at high \pT suggests that the fragmentation into
light mesons is the same in all collisions systems.
At $\pT<2$~\GeVc, the \e/\pai ratio shows a clear pattern of deviation from the ratio predicted by 
the \mT scaling, confirming a \mT scaling violation observed earlier in \pA collisions at \sNNSPS and in pp 
collisions at \sL and \sLL.
The presence of radial flow 
effects in small systems and contributions from heavier-mesons decays to the \e and \pai production spectra
are among possible interpretations of the \mT scaling violation.
The comparison to different model calculations suggests that hydrodynamical flow may help to describe the measured spectra at low \pT. 
Theoretical calculations using DPMJET and HIJING are very close to the \mT scaling prediction and therefore 
overestimate the measured ratio. The \e/\pai ratio is reproduced in the complete \pT range by the VISHNU calculations 
although any conclusions above 1.5 \GeVc are difficult to extract as the spectra
were underestimated by large factors.
For \pT $<$ 3~\GeVc, the \e/\pai ratio calculated by EPOS3 is closer to the measured data than the \mT scaling prediction, and it agrees
with the data in the intermediate \pT range 2 $<$ \pT $<$ 5~\GeVc.
These model comparisons support the interpretation that radial flow plays a role in collisions of small systems at the LHC.

The measured nuclear modification factors \RpPb for the \pai and \e meson are
consistent with unity at $\pT>2$~\GeVc which confirms previously
reported measurements at RHIC  \cite{Adler:2006wg,Adams:2006nd} and LHC 
\cite{ALICE:2012mj,Chatrchyan:2013eya,Abelev:2013haa,Aad:2016zif,Khachatryan:2016odn}.  
Theoretical calculations based on the latest nPDFs and a model based on the CGC framework are able to describe \RpPb well.
These results support the interpretation that the neutral pion suppression in central \Pb collisions is due to 
parton energy loss in the hot QCD medium.

These data are an important input for theoretical models aiming at the description of particle 
production in small systems at LHC energies and provide additional constraints on nPDFs and identified FFs.

\newenvironment{acknowledgement}{\relax}{\relax}
\begin{acknowledgement}
\section*{Acknowledgements}
The ALICE Collaboration would like to thank H.~M\"antysaari and T.~Lappi for providing the CGC theory 
calculations, I.~Helenius and W. Vogelsang for providing the pQCD calculations with EPPS16 and nCTEQ 
respectively, and C. Shen for providing the hydrodynamical calculations. We also would like to thank 
K. Werner for helpful discussions.

The ALICE Collaboration would like to thank all its engineers and technicians for their invaluable contributions to the construction of the experiment and the CERN accelerator teams for the outstanding performance of the LHC complex.
The ALICE Collaboration gratefully acknowledges the resources and support provided by all Grid centres and the Worldwide LHC Computing Grid (WLCG) collaboration.
The ALICE Collaboration acknowledges the following funding agencies for their support in building and running the ALICE detector:
A. I. Alikhanyan National Science Laboratory (Yerevan Physics Institute) Foundation (ANSL), State Committee of Science and World Federation of Scientists (WFS), Armenia;
Austrian Academy of Sciences and Nationalstiftung f\"{u}r Forschung, Technologie und Entwicklung, Austria;
Ministry of Communications and High Technologies, National Nuclear Research Center, Azerbaijan;
Conselho Nacional de Desenvolvimento Cient\'{\i}fico e Tecnol\'{o}gico (CNPq), Universidade Federal do Rio Grande do Sul (UFRGS), Financiadora de Estudos e Projetos (Finep) and Funda\c{c}\~{a}o de Amparo \`{a} Pesquisa do Estado de S\~{a}o Paulo (FAPESP), Brazil;
Ministry of Science \& Technology of China (MSTC), National Natural Science Foundation of China (NSFC) and Ministry of Education of China (MOEC) , China;
Ministry of Science, Education and Sport and Croatian Science Foundation, Croatia;
Ministry of Education, Youth and Sports of the Czech Republic, Czech Republic;
The Danish Council for Independent Research | Natural Sciences, the Carlsberg Foundation and Danish National Research Foundation (DNRF), Denmark;
Helsinki Institute of Physics (HIP), Finland;
Commissariat \`{a} l'Energie Atomique (CEA) and Institut National de Physique Nucl\'{e}aire et de Physique des Particules (IN2P3) and Centre National de la Recherche Scientifique (CNRS), France;
Bundesministerium f\"{u}r Bildung, Wissenschaft, Forschung und Technologie (BMBF) and GSI Helmholtzzentrum f\"{u}r Schwerionenforschung GmbH, Germany;
General Secretariat for Research and Technology, Ministry of Education, Research and Religions, Greece;
National Research, Development and Innovation Office, Hungary;
Department of Atomic Energy Government of India (DAE), Department of Science and Technology, Government of India (DST), University Grants Commission, Government of India (UGC) and Council of Scientific and Industrial Research (CSIR), India;
Indonesian Institute of Science, Indonesia;
Centro Fermi - Museo Storico della Fisica e Centro Studi e Ricerche Enrico Fermi and Istituto Nazionale di Fisica Nucleare (INFN), Italy;
Institute for Innovative Science and Technology , Nagasaki Institute of Applied Science (IIST), Japan Society for the Promotion of Science (JSPS) KAKENHI and Japanese Ministry of Education, Culture, Sports, Science and Technology (MEXT), Japan;
Consejo Nacional de Ciencia (CONACYT) y Tecnolog\'{i}a, through Fondo de Cooperaci\'{o}n Internacional en Ciencia y Tecnolog\'{i}a (FONCICYT) and Direcci\'{o}n General de Asuntos del Personal Academico (DGAPA), Mexico;
Nederlandse Organisatie voor Wetenschappelijk Onderzoek (NWO), Netherlands;
The Research Council of Norway, Norway;
Commission on Science and Technology for Sustainable Development in the South (COMSATS), Pakistan;
Pontificia Universidad Cat\'{o}lica del Per\'{u}, Peru;
Ministry of Science and Higher Education and National Science Centre, Poland;
Korea Institute of Science and Technology Information and National Research Foundation of Korea (NRF), Republic of Korea;
Ministry of Education and Scientific Research, Institute of Atomic Physics and Romanian National Agency for Science, Technology and Innovation, Romania;
Joint Institute for Nuclear Research (JINR), Ministry of Education and Science of the Russian Federation and National Research Centre Kurchatov Institute, Russia;
Ministry of Education, Science, Research and Sport of the Slovak Republic, Slovakia;
National Research Foundation of South Africa, South Africa;
Centro de Aplicaciones Tecnol\'{o}gicas y Desarrollo Nuclear (CEADEN), Cubaenerg\'{\i}a, Cuba and Centro de Investigaciones Energ\'{e}ticas, Medioambientales y Tecnol\'{o}gicas (CIEMAT), Spain;
Swedish Research Council (VR) and Knut \& Alice Wallenberg Foundation (KAW), Sweden;
European Organization for Nuclear Research, Switzerland;
National Science and Technology Development Agency (NSDTA), Suranaree University of Technology (SUT) and Office of the Higher Education Commission under NRU project of Thailand, Thailand;
Turkish Atomic Energy Agency (TAEK), Turkey;
National Academy of  Sciences of Ukraine, Ukraine;
Science and Technology Facilities Council (STFC), United Kingdom;
National Science Foundation of the United States of America (NSF) and United States Department of Energy, Office of Nuclear Physics (DOE NP), United States of America.
\end{acknowledgement}


\bibliographystyle{utphys}   
\bibliography{biblio.bib}

\appendix

\section{Parameters of TCM fits}
The parameters of the two-component model fits to the reference \pai and \e meson spectra in pp 
collisions at $\sqrt{s}=5.02$~TeV shown in 
\hyperref[fig:Pi0EtaYields]{Fig.~\ref*{fig:Pi0EtaYields}} are given in the
\hyperref[tab:FitParametersPPref]{Table~\ref*{tab:FitParametersPPref}}. 
The \pai and \e meson references were calculated using \hyperref[eq:RpPb]{Eq.~\ref*{eq:RpPb}}, 
from the combined spectra in p-Pb collisions (\hyperref[fig:Pi0EtaYields]{Fig.~\ref*{fig:Pi0EtaYields}}), 
and combined \RpPb (\hyperref[fig:RpPb]{Fig.~\ref*{fig:RpPb}}).
\begin{table}[pht]
\small
\centering 
\renewcommand{\arraystretch}{1.15}
  \begin{tabular}{p{1.8cm}|c|c|}
  \cline{2-3}
                 & \pai spectrum fit             & \e spectrum fit		\\ \hline
  \multicolumn{1}{|p{2.8cm}|}{  $A_e$ (pb GeV$^{-2}c^3$)}           & 3.76 $\times$ 10$^{11}$ & 5.75 $\times$ 10$^{9}$ 	\\ 
   \multicolumn{1}{|p{2.8cm}|}{  $T_e$ (GeV/$c$)} & 0.151 & 0.252 	\\ 
  \multicolumn{1}{|p{2.8cm}|}{  $A$ (pb GeV$^{-2}c^3$)}           & 3.1 $\times$ 10$^{10}$ & 1.21 $\times$ 10$^{9}$ 	\\ 
   \multicolumn{1}{|p{2.8cm}|}{  $T$ (GeV/$c$)} & 0.585 & 0.916 	\\ 
 \multicolumn{1}{|p{2.8cm}|}{  $n$ }          & 3.09   & 3.12	\\ \hline 		 
  \end{tabular}
  \caption{Fit parameters of the TCM fits to the reference \pai and \e spectra in pp collisions at $\sqrt{s}=5.02$~TeV.}
  \label{tab:FitParametersPPref}
\end{table}

\newpage
\section{The ALICE Collaboration}
\label{app:collab}

\begingroup
\small
\begin{flushleft}
S.~Acharya\Irefn{org137}\And 
D.~Adamov\'{a}\Irefn{org93}\And 
J.~Adolfsson\Irefn{org80}\And 
M.M.~Aggarwal\Irefn{org97}\And 
G.~Aglieri Rinella\Irefn{org35}\And 
M.~Agnello\Irefn{org32}\And 
N.~Agrawal\Irefn{org47}\And 
Z.~Ahammed\Irefn{org137}\And 
S.U.~Ahn\Irefn{org76}\And 
S.~Aiola\Irefn{org142}
% ...
\And
\renewcommand\labelenumi{\textsuperscript{\theenumi}~}

\section*{Affiliation notes}
\renewcommand\theenumi{\roman{enumi}}
\begin{Authlist}
\item \Adef{org*}Deceased
\item \Adef{orgI}Dipartimento DET del Politecnico di Torino, Turin, Italy
\item \Adef{orgII}M.V. Lomonosov Moscow State University, D.V. Skobeltsyn Institute of Nuclear, Physics, Moscow, Russia
\item \Adef{orgIII}Department of Applied Physics, Aligarh Muslim University, Aligarh, India
\item \Adef{orgIV}Institute of Theoretical Physics, University of Wroclaw, Poland
\end{Authlist}

\section*{Collaboration Institutes}
\renewcommand\theenumi{\arabic{enumi}~}
\begin{Authlist}
\item \Idef{org1}A.I. Alikhanyan National Science Laboratory (Yerevan Physics Institute) Foundation, Yerevan, Armenia
\item \Idef{org2}Benem\'{e}rita Universidad Aut\'{o}noma de Puebla, Puebla, Mexico
\item \Idef{org3}Bogolyubov Institute for Theoretical Physics, National Academy of Sciences of Ukraine, Kiev, Ukraine
\item \Idef{org4}Bose Institute, Department of Physics  and Centre for Astroparticle Physics and Space Science (CAPSS), Kolkata, India
\item \Idef{org5}Budker Institute for Nuclear Physics, Novosibirsk, Russia
\item \Idef{org6}California Polytechnic State University, San Luis Obispo, California, United States
\item \Idef{org7}Central China Normal University, Wuhan, China
\item \Idef{org8}Centre de Calcul de l'IN2P3, Villeurbanne, Lyon, France
\item \Idef{org9}Centro de Aplicaciones Tecnol\'{o}gicas y Desarrollo Nuclear (CEADEN), Havana, Cuba
\item \Idef{org10}Centro de Investigaci\'{o}n y de Estudios Avanzados (CINVESTAV), Mexico City and M\'{e}rida, Mexico
\item \Idef{org11}Centro Fermi - Museo Storico della Fisica e Centro Studi e Ricerche ``Enrico Fermi', Rome, Italy
\item \Idef{org12}Chicago State University, Chicago, Illinois, United States
\item \Idef{org13}China Institute of Atomic Energy, Beijing, China
\item \Idef{org14}Chonbuk National University, Jeonju, Republic of Korea
\item \Idef{org15}Comenius University Bratislava, Faculty of Mathematics, Physics and Informatics, Bratislava, Slovakia
\item \Idef{org16}COMSATS Institute of Information Technology (CIIT), Islamabad, Pakistan
\item \Idef{org17}Creighton University, Omaha, Nebraska, United States
\item \Idef{org18}Department of Physics, Aligarh Muslim University, Aligarh, India
\item \Idef{org19}Department of Physics, Ohio State University, Columbus, Ohio, United States
\item \Idef{org20}Department of Physics, Pusan National University, Pusan, Republic of Korea
\item \Idef{org21}Department of Physics, Sejong University, Seoul, Republic of Korea
\item \Idef{org22}Department of Physics, University of Oslo, Oslo, Norway
\item \Idef{org23}Department of Physics and Technology, University of Bergen, Bergen, Norway
\item \Idef{org24}Dipartimento di Fisica dell'Universit\`{a} 'La Sapienza' and Sezione INFN, Rome, Italy
\item \Idef{org25}Dipartimento di Fisica dell'Universit\`{a} and Sezione INFN, Cagliari, Italy
\item \Idef{org26}Dipartimento di Fisica dell'Universit\`{a} and Sezione INFN, Trieste, Italy
\item \Idef{org27}Dipartimento di Fisica dell'Universit\`{a} and Sezione INFN, Turin, Italy
\item \Idef{org28}Dipartimento di Fisica e Astronomia dell'Universit\`{a} and Sezione INFN, Bologna, Italy
\item \Idef{org29}Dipartimento di Fisica e Astronomia dell'Universit\`{a} and Sezione INFN, Catania, Italy
\item \Idef{org30}Dipartimento di Fisica e Astronomia dell'Universit\`{a} and Sezione INFN, Padova, Italy
\item \Idef{org31}Dipartimento di Fisica `E.R.~Caianiello' dell'Universit\`{a} and Gruppo Collegato INFN, Salerno, Italy
\item \Idef{org32}Dipartimento DISAT del Politecnico and Sezione INFN, Turin, Italy
\item \Idef{org33}Dipartimento di Scienze e Innovazione Tecnologica dell'Universit\`{a} del Piemonte Orientale and INFN Sezione di Torino, Alessandria, Italy
\item \Idef{org34}Dipartimento Interateneo di Fisica `M.~Merlin' and Sezione INFN, Bari, Italy
\item \Idef{org35}European Organization for Nuclear Research (CERN), Geneva, Switzerland
\item \Idef{org36}Faculty of Engineering and Business Administration, Western Norway University of Applied Sciences, Bergen, Norway
\item \Idef{org37}Faculty of Nuclear Sciences and Physical Engineering, Czech Technical University in Prague, Prague, Czech Republic
\item \Idef{org38}Faculty of Science, P.J.~\v{S}af\'{a}rik University, Ko\v{s}ice, Slovakia
\item \Idef{org39}Frankfurt Institute for Advanced Studies, Johann Wolfgang Goethe-Universit\"{a}t Frankfurt, Frankfurt, Germany
\item \Idef{org40}Gangneung-Wonju National University, Gangneung, Republic of Korea
\item \Idef{org41}Gauhati University, Department of Physics, Guwahati, India
\item \Idef{org42}Helmholtz-Institut f\"{u}r Strahlen- und Kernphysik, Rheinische Friedrich-Wilhelms-Universit\"{a}t Bonn, Bonn, Germany
\item \Idef{org43}Helsinki Institute of Physics (HIP), Helsinki, Finland
\item \Idef{org44}Hiroshima University, Hiroshima, Japan
\item \Idef{org45}Hochschule Worms, Zentrum  f\"{u}r Technologietransfer und Telekommunikation (ZTT), Worms, Germany
\item \Idef{org46}Horia Hulubei National Institute of Physics and Nuclear Engineering, Bucharest, Romania
\item \Idef{org47}Indian Institute of Technology Bombay (IIT), Mumbai, India
\item \Idef{org48}Indian Institute of Technology Indore, Indore, India
\item \Idef{org49}Indonesian Institute of Sciences, Jakarta, Indonesia
\item \Idef{org50}INFN, Laboratori Nazionali di Frascati, Frascati, Italy
\item \Idef{org51}INFN, Sezione di Bari, Bari, Italy
\item \Idef{org52}INFN, Sezione di Bologna, Bologna, Italy
\item \Idef{org53}INFN, Sezione di Cagliari, Cagliari, Italy
\item \Idef{org54}INFN, Sezione di Catania, Catania, Italy
\item \Idef{org55}INFN, Sezione di Padova, Padova, Italy
\item \Idef{org56}INFN, Sezione di Roma, Rome, Italy
\item \Idef{org57}INFN, Sezione di Torino, Turin, Italy
\item \Idef{org58}INFN, Sezione di Trieste, Trieste, Italy
\item \Idef{org59}Inha University, Incheon, Republic of Korea
\item \Idef{org60}Institut de Physique Nucl\'{e}aire d'Orsay (IPNO), Institut National de Physique Nucl\'{e}aire et de Physique des Particules (IN2P3/CNRS), Universit\'{e} de Paris-Sud, Universit\'{e} Paris-Saclay, Orsay, France
\item \Idef{org61}Institute for Nuclear Research, Academy of Sciences, Moscow, Russia
\item \Idef{org62}Institute for Subatomic Physics of Utrecht University, Utrecht, Netherlands
\item \Idef{org63}Institute for Theoretical and Experimental Physics, Moscow, Russia
\item \Idef{org64}Institute of Experimental Physics, Slovak Academy of Sciences, Ko\v{s}ice, Slovakia
\item \Idef{org65}Institute of Physics, Bhubaneswar, India
\item \Idef{org66}Institute of Physics of the Czech Academy of Sciences, Prague, Czech Republic
\item \Idef{org67}Institute of Space Science (ISS), Bucharest, Romania
\item \Idef{org68}Institut f\"{u}r Kernphysik, Johann Wolfgang Goethe-Universit\"{a}t Frankfurt, Frankfurt, Germany
\item \Idef{org69}Instituto de Ciencias Nucleares, Universidad Nacional Aut\'{o}noma de M\'{e}xico, Mexico City, Mexico
\item \Idef{org70}Instituto de F\'{i}sica, Universidade Federal do Rio Grande do Sul (UFRGS), Porto Alegre, Brazil
\item \Idef{org71}Instituto de F\'{\i}sica, Universidad Nacional Aut\'{o}noma de M\'{e}xico, Mexico City, Mexico
\item \Idef{org72}iThemba LABS, National Research Foundation, Somerset West, South Africa
\item \Idef{org73}Johann-Wolfgang-Goethe Universit\"{a}t Frankfurt Institut f\"{u}r Informatik, Fachbereich Informatik und Mathematik, Frankfurt, Germany
\item \Idef{org74}Joint Institute for Nuclear Research (JINR), Dubna, Russia
\item \Idef{org75}Konkuk University, Seoul, Republic of Korea
\item \Idef{org76}Korea Institute of Science and Technology Information, Daejeon, Republic of Korea
\item \Idef{org77}KTO Karatay University, Konya, Turkey
\item \Idef{org78}Laboratoire de Physique Subatomique et de Cosmologie, Universit\'{e} Grenoble-Alpes, CNRS-IN2P3, Grenoble, France
\item \Idef{org79}Lawrence Berkeley National Laboratory, Berkeley, California, United States
\item \Idef{org80}Lund University Department of Physics, Division of Particle Physics, Lund, Sweden
\item \Idef{org81}Nagasaki Institute of Applied Science, Nagasaki, Japan
\item \Idef{org82}Nara Women{'}s University (NWU), Nara, Japan
\item \Idef{org83}National and Kapodistrian University of Athens, School of Science, Department of Physics , Athens, Greece
\item \Idef{org84}National Centre for Nuclear Research, Warsaw, Poland
\item \Idef{org85}National Institute of Science Education and Research, HBNI, Jatni, India
\item \Idef{org86}National Nuclear Research Center, Baku, Azerbaijan
\item \Idef{org87}National Research Centre Kurchatov Institute, Moscow, Russia
\item \Idef{org88}Niels Bohr Institute, University of Copenhagen, Copenhagen, Denmark
\item \Idef{org89}Nikhef, National institute for subatomic physics, Amsterdam, Netherlands
\item \Idef{org90}NRC ¿Kurchatov Institute¿ ¿ IHEP , Protvino, Russia
\item \Idef{org91}NRNU Moscow Engineering Physics Institute, Moscow, Russia
\item \Idef{org92}Nuclear Physics Group, STFC Daresbury Laboratory, Daresbury, United Kingdom
\item \Idef{org93}Nuclear Physics Institute of the Czech Academy of Sciences, \v{R}e\v{z} u Prahy, Czech Republic
\item \Idef{org94}Oak Ridge National Laboratory, Oak Ridge, Tennessee, United States
\item \Idef{org95}Petersburg Nuclear Physics Institute, Gatchina, Russia
\item \Idef{org96}Physics department, Faculty of science, University of Zagreb, Zagreb, Croatia
\item \Idef{org97}Physics Department, Panjab University, Chandigarh, India
\item \Idef{org98}Physics Department, University of Jammu, Jammu, India
\item \Idef{org99}Physics Department, University of Rajasthan, Jaipur, India
\item \Idef{org100}Physikalisches Institut, Eberhard-Karls-Universit\"{a}t T\"{u}bingen, T\"{u}bingen, Germany
\item \Idef{org101}Physikalisches Institut, Ruprecht-Karls-Universit\"{a}t Heidelberg, Heidelberg, Germany
\item \Idef{org102}Physik Department, Technische Universit\"{a}t M\"{u}nchen, Munich, Germany
\item \Idef{org103}Research Division and ExtreMe Matter Institute EMMI, GSI Helmholtzzentrum f\"ur Schwerionenforschung GmbH, Darmstadt, Germany
\item \Idef{org104}Rudjer Bo\v{s}kovi\'{c} Institute, Zagreb, Croatia
\item \Idef{org105}Russian Federal Nuclear Center (VNIIEF), Sarov, Russia
\item \Idef{org106}Saha Institute of Nuclear Physics, Kolkata, India
\item \Idef{org107}School of Physics and Astronomy, University of Birmingham, Birmingham, United Kingdom
\item \Idef{org108}Secci\'{o}n F\'{\i}sica, Departamento de Ciencias, Pontificia Universidad Cat\'{o}lica del Per\'{u}, Lima, Peru
\item \Idef{org109}Stefan Meyer Institut f\"{u}r Subatomare Physik (SMI), Vienna, Austria
\item \Idef{org110}SUBATECH, IMT Atlantique, Universit\'{e} de Nantes, CNRS-IN2P3, Nantes, France
\item \Idef{org111}Suranaree University of Technology, Nakhon Ratchasima, Thailand
\item \Idef{org112}Technical University of Ko\v{s}ice, Ko\v{s}ice, Slovakia
\item \Idef{org113}Technische Universit\"{a}t M\"{u}nchen, Excellence Cluster 'Universe', Munich, Germany
\item \Idef{org114}The Henryk Niewodniczanski Institute of Nuclear Physics, Polish Academy of Sciences, Cracow, Poland
\item \Idef{org115}The University of Texas at Austin, Austin, Texas, United States
\item \Idef{org116}Universidad Aut\'{o}noma de Sinaloa, Culiac\'{a}n, Mexico
\item \Idef{org117}Universidade de S\~{a}o Paulo (USP), S\~{a}o Paulo, Brazil
\item \Idef{org118}Universidade Estadual de Campinas (UNICAMP), Campinas, Brazil
\item \Idef{org119}Universidade Federal do ABC, Santo Andre, Brazil
\item \Idef{org120}University College of Southeast Norway, Tonsberg, Norway
\item \Idef{org121}University of Cape Town, Cape Town, South Africa
\item \Idef{org122}University of Houston, Houston, Texas, United States
\item \Idef{org123}University of Jyv\"{a}skyl\"{a}, Jyv\"{a}skyl\"{a}, Finland
\item \Idef{org124}University of Liverpool, Liverpool, United Kingdom
\item \Idef{org125}University of Split, Faculty of Electrical Engineering, Mechanical Engineering and Naval Architecture, Split, Croatia
\item \Idef{org126}University of Tennessee, Knoxville, Tennessee, United States
\item \Idef{org127}University of the Witwatersrand, Johannesburg, South Africa
\item \Idef{org128}University of Tokyo, Tokyo, Japan
\item \Idef{org129}University of Tsukuba, Tsukuba, Japan
\item \Idef{org130}Universit\'{e} Clermont Auvergne, CNRS/IN2P3, LPC, Clermont-Ferrand, France
\item \Idef{org131}Universit\'{e} de Lyon, Universit\'{e} Lyon 1, CNRS/IN2P3, IPN-Lyon, Villeurbanne, Lyon, France
\item \Idef{org132}Universit\'{e} de Strasbourg, CNRS, IPHC UMR 7178, F-67000 Strasbourg, France, Strasbourg, France
\item \Idef{org133} Universit\'{e} Paris-Saclay Centre d¿\'Etudes de Saclay (CEA), IRFU, Department de Physique Nucl\'{e}aire (DPhN), Saclay, France
\item \Idef{org134}Universit\`{a} degli Studi di Pavia, Pavia, Italy
\item \Idef{org135}Universit\`{a} di Brescia, Brescia, Italy
\item \Idef{org136}V.~Fock Institute for Physics, St. Petersburg State University, St. Petersburg, Russia
\item \Idef{org137}Variable Energy Cyclotron Centre, Kolkata, India
\item \Idef{org138}Warsaw University of Technology, Warsaw, Poland
\item \Idef{org139}Wayne State University, Detroit, Michigan, United States
\item \Idef{org140}Westf\"{a}lische Wilhelms-Universit\"{a}t M\"{u}nster, Institut f\"{u}r Kernphysik, M\"{u}nster, Germany
\item \Idef{org141}Wigner Research Centre for Physics, Hungarian Academy of Sciences, Budapest, Hungary
\item \Idef{org142}Yale University, New Haven, Connecticut, United States
\item \Idef{org143}Yonsei University, Seoul, Republic of Korea
\end{Authlist}
\endgroup
\end{document}